\documentclass[twocolumn,twocolappendix]{aastex63}

\usepackage{ragged2e}
\usepackage{amsmath}
\usepackage[nameinlink]{cleveref}
\usepackage[]{natbib}
\setcitestyle{comma}
\usepackage{lineno}[mathlines]
\usepackage[utf8]{inputenc}
\usepackage[T1]{fontenc}

\makeatletter
\newcommand*{\linktocite}[2]{%
  \hyper@natlinkstart{#1}#2\hyper@natlinkend}
\makeatother

\newcommand{\Eq}[1]{Equation\,(\ref{#1})}

\newcommand{\Sec}[1]{Section~\ref{#1}}

\newcommand{\Fig}[1]{Figure~\ref{#1}}

\def\K{\; {\rm K}}

\shorttitle{Limited hysteresis in hot Jupiter circulation}
\shortauthors{T.D. Komacek}
\graphicspath{{./}{figs/}}
\begin{document}

\title{Limited hysteresis in the atmospheric dynamics of hot Jupiters}

\correspondingauthor{Thaddeus D. Komacek}
\email{tad.komacek@physics.ox.ac.uk}
\author[0000-0002-9258-5311]{Thaddeus D. Komacek}
\affiliation{Department of Physics (Atmospheric, Oceanic and Planetary Physics), University of Oxford, Oxford OX1 3PU, UK}
\affiliation{Department of Astronomy, University of Maryland, College Park, MD 20742, USA}




\begin{abstract}
Over the past two decades, a coherent picture has emerged of the atmospheric dynamics of hot Jupiters from a combination of three-dimensional general circulation models (GCMs) and astronomical observations. This paradigm consists of hot Jupiters being spin-synchronized due to their close-in orbit, with a resulting large day-to-night irradiation gradient driving a day-to-night temperature contrast. This day-to-night temperature contrast in turn raises day-to-night pressure gradients that are balanced by a circulation with wind speeds on the order of km~s$^{-1}$. The dominant feature of this circulation is a superrotating equatorial jet, maintained by eddy-mean flow interactions that pump momentum into the jet.
%from higher latitudes toward the equator. 
In this work, I explore the dependence of this circulation paradigm on the initial thermal and dynamical conditions in atmospheric circulation models of  hot Jupiters. To do so, I conduct MITgcm simulations of the atmospheric circulation of hot Jupiters with both varying initial wind directions and initial temperature profiles. I find that the results are insensitive to the initial conditions, implying that the current paradigm of hot Jupiter circulation exhibits at most limited hysteresis. I demonstrate that there is a single characteristic wind speed of hot Jupiters for given planetary and atmospheric parameters
%is no bifurcation between superrotating and subrotating states for hot Jupiters 
using an idealized scaling theory, and discuss implications for 
%phase curve and eclipse map 
the interpretation of hot Jupiter observations. 
%Notes on searching for bifurcations in HJ climate analogous to \citep{sergeev:2022ab} for tidally locked rocky exoplanets. 
\end{abstract}



\section{Introduction}
Simulations of the atmospheric dynamics of hot Jupiters over the past twenty-two years since \cite{showman_2002} broadly agree in the key features of their circulation. This standard model of hot Jupiter atmospheric circulation includes a superrotating eastward equatorial jet that induces a photospheric eastward hot spot offset and large day-to-night temperature contrasts driven by strong radiative cooling in these hot atmospheres \citep{Heng:2014b,Showman:2020rev}. Though models agree in broad qualitative predictions, inter-model differences exist, especially with regard to the amplitude of variability and potential sensitivity to initial conditions \citep{Liu:2013,Cho:2015,Cho:2021wb,2022MNRAS.511.3584S}. 

The potential for hysteresis (i.e., a dependence of the resulting circulation on initial conditions) is critical to explore in order to determine whether the basic dynamical state of hot Jupiters is robust. Notably, bifurcations in the jet structure depending on the initial conditions have been found in GCMs of tidally locked temperate terrestrial planets \citep{sergeev:2022ab}, the atmospheric circulation in simulations of sub-Neptunes with deep atmospheres display bistability \citep{Wang:2020aa}, and the deep atmosphere has been demonstrated to potentially impact the direction of the equatorial jet on hot Jupiters \citep{Carone:2019aa} -- as a result, 
%However, hot Jupiters lie in a much stronger day-to-night forcing regime than temperate and warm terrestrial planets and sub-Neptunes, which may limit their potential for hysteresis and time-variability   but 
there is a need to determine if bistability may also occur in the atmospheric dynamics of hot Jupiters. 
%due to their strong forcing. 
%Though previous work has disagreed on the sensitivity of near-photospheric hot Jupiter circulation to initial conditions \citep{Liu:2013,Cho:2015}, recent deep simulations of the ultra-hot Jupiter WASP-76b have found that though evolution to the final temperature profile is slow due to long radiative and advective timescales at depth, the deep thermal structure is independent of the initial temperature profile \citep{Sainsbury-Martinez:2023aa}. This may imply that the atmospheric circulation of hot Jupiters is distinct from that of other planets with thick atmospheres 

Determining the extent to which hot Jupiter dynamics can undergo hysteresis is critical for the interpretation of past, current and future observations with 3D models. This is especially important for JWST-quality secondary eclipse observations \citep{Coulombe:2023aa} and phase curves \citep{Mikal-Evans:2023aa,Bell:2024aa}, as our observational sensitivity is now sufficient to constrain the 3D thermal structure of hot Jupiters \citep{Mansfield:2020ab,Challener:2022,Hammond:2024aa}. Notably, the baseline expectation from the standard model of hot Jupiter atmospheric dynamics is for equatorial superrotation and resulting eastward hot spot offsets near the thermal photosphere \citep{Roth:2024aa}, with day-to-night flow dominating at lower pressures, especially for hotter planets \citep{Kempton:2012vk,showman_2013_doppler,Zhang:2020rev}. However, observations do not necessarily find eastward hot spot offsets to be ubiquitous \citep{Dang:2018aa,Zhang:2017a,Bell:2021aa,2022AJ....163..256M}. Proposed explanations for westward hot spot offsets include magnetic effects \citep{Rogers:2017}, non-synchronous rotation \citep{Rauscher:2014aa}, and temperature structures being dominated by a baroclinic wave structure with height maxima that are shifted west of the substellar point near the thermal photosphere
\citep{Lewis:2022aa}. One additional potential explanation that has not been explored in detail for hot Jupiters could be a bifurcation in the jet structure resulting in weak equatorial flow and strong mid-latitude flow, as found for temperate rocky planets \citep{sergeev:2022ab}. 

Previous work by \cite{Shell:2004} has shown an abrupt emergence of superrotation with increasing prescribed momentum forcing strength in axisymmetric idealized models. The \cite{Shell:2004} model may provide an explanation for the bifurcations found in previous tidally locked planet simulations of \cite{sergeev:2022ab}, where the strength of superrotation depends on the initial conditions as well as convective and cloud schemes. As in \cite{Shell:2004}, there is a need to include a vertical momentum transport term in the model setup in order to reproduce superrotation in analytic and numerical shallow-water hot Jupiter models \citep{Showman:2010,Showman_Polvani_2011}. Similarly, three-dimensional studies of the mechanisms governing superrotation in hot Jupiters require both horizontal and vertical eddy momentum transport by planetary-scale waves to drive superrotating flow \citep{Tsai:2014,Hammond:2018aa}. 

It is plausible that because the day-to-night radiative forcing regime and thus resulting planetary-scale wave pattern of hot Jupiters is different from those of terrestrial planets there could be a significant reduction in the magnitude of any possible dynamical bifurcation. Lastly, note that the strong day-to-night forcing of hot Jupiters also leads to numerical limitations in the predictions of equatorial jet speeds due to the potential for over-damping in simulations \citep{Koll:2017,2021MNRAS.504.5172S,Hammond:2022aa}, requiring further exploration of the impacts of both explicit and numerical dissipation on hot Jupiter atmospheric dynamics. 

In this work I study the potential for hysteresis in the atmospheric dynamics of hot Jupiters near dynamical regime transitions, as has been found previously in simulations of tidally locked temperate terrestrial planets \citep{sergeev:2022ab}. I specifically explore the potential for hysteresis at the dynamical regime transitions where the length scale at which turbulence manifests into jets (i.e., the Rhines scale) and the length scale on which gravity waves are affected by rotation (i.e., the Rossby deformation radius) are comparable to the planetary radius \citep{Haqq2018}. Notably, the atmospheric circulation of hot Jupiters is more strongly forced by irradiation than that of temperate terrestrial planets. This warrants a theoretical comparison between the dynamical states of these objects as well as suites of numerical experiments in order to determine whether a bifurcation in the strength of superrotation can manifest similarly at the boundaries of planetary dynamical regimes. 

This work is outlined as follows. First, I derive the parameter regimes in which hot Jupiters may undergo transitions between dynamical states where the Rhines scale or Rossby deformation radius is larger or smaller than the planetary radius, and connect these to previously established dynamical regimes for tidally locked planets (\Sec{sec:dynregime}). Then, I detail the numerical setup of and present results from two suites of GCM simulations that study the potential for dynamical regime transitions (\Sec{sec:methods} and \Sec{sec:results}). I then interpret these results by developing an idealized scaling theory for the dependence of the equatorial jet speed on momentum forcing (\Sec{sec:theory}). I finally discuss limitations, future work, and an outlook (\Sec{sec:disc}), and state key conclusions (\Sec{sec:conc}). 
%\vspace{-1.1cm}
\section{Potential dynamical regime transitions}
\label{sec:dynregime}
\subsection{Rhines scale and Rossby number}
%\cite{Haqq2018} define the transition between slow and Rhines rotators to occur where the Rhines scale $\lambda_\beta = a$, where $a$ is the planetary radius. 
%n.b. if you include a factor of pi in rhines scale then T_eq,beta increases by a factor of pi^(1/3) and becomes ~1900 K for typical values. Might be worth running a larger suite of GCMs.
The Rhines scale is $\lambda_\beta \sim \sqrt{U/\beta}$ \citep{Rhines:1975aa}, where $U$ is a characteristic wind speed (commonly taken to be the root-mean-squre, RMS, wind speed as in \citealp{Rhines:1975aa}) and $\beta = 2 \Omega\mathrm{cos}(\phi)/a$ where $\Omega$ is the planetary rotation rate, $\phi$ is latitude, and $a$ is the planetary radius. Hence, one can express that the Rhines scale is approximately equal to the planetary radius when\footnote{Note that these results depend on the exact definition of the Rhines scale and relevant planetary scale. For instance, one may also define $\lambda_\beta = \pi \sqrt{U/\beta}$, and compare it to the equator-to-pole distance equal to one-quarter of the spherical planetary circumference $\pi a/2$, leading to $\sqrt{\frac{2Ua}{\Omega \mathrm{cos}(\phi)}} \approx a$ rather than \Eq{eq:uabalancerhines}. In this work, we follow \cite{Haqq2018} and provide relationships comparing approximate dynamical length scales to the planetary radius.}:
\begin{equation}
    \label{eq:uabalancerhines}
    \sqrt{\frac{U}{\beta}} = \sqrt{\frac{Ua}{2\Omega \mathrm{cos}(\phi)}} \approx a ~\mathrm{.}
\end{equation}
One can then calculate the rotation period $P_{\beta} = 2\pi/\Omega$ at this transition at the equator (i.e., $\phi = 0^\circ$) as a function of planetary radius and wind speed:
\begin{equation}
    P_\beta \approx \frac{4 \pi a}{U}~\mathrm{.} 
    %would just be \pi a / U if using 1/4 circumference
\end{equation}
I then assume that the hot Jupiter is tidally locked, with a rotation period equal to its orbital period. I use Kepler's third law and the inverse-square law to write the orbital period $P_\mathrm{orb}$ as a function of stellar luminosity $L_\star$, stellar mass $M_\star$, and incident stellar flux $F_\star$
\begin{equation}
    \label{eq:kepler}
    P_\mathrm{orb} = \frac{\pi^{1/4}}{\sqrt{2GM_\star}} \left(\frac{L_\star}{F_\star}\right)^{3/4}~\mathrm{.}
\end{equation}
Equating $P_\beta = P_\mathrm{orb}$ and substituting the full-redistribution zero-albedo equilibrium temperature $T_\mathrm{eq} = [F_\star/(4\sigma)]^{1/4}$, I find the dependence of the transition equilibrium temperature where the Rhines scale equals the planetary radius as a function of stellar, planetary, and atmospheric parameters:
\begin{equation}
\label{eq:teqbeta}
\begin{split}
    T_{eq,\beta} \approx & 1300~\mathrm{K}  \left(\frac{M_\star}{M_\odot}\right)^{-1/6} \left(\frac{L_\star}{L_\odot}\right)^{1/4} \\  & \times \left(\frac{a}{1.38~R_\mathrm{{Jup,eq}}}\right)^{-1/3}\left(\frac{U}{4~\mathrm{km}~\mathrm{s}^{-1}}\right)^{1/3}~\mathrm{.}
\end{split}
\end{equation}
All stellar values in \Eq{eq:teqbeta} are normalized to the Solar value, the planetary radius is normalized to that of HD 209458b, and the wind speed is normalized to that in a typical hot Jupiter simulation for HD 209458b-like parameters \citep{Komacek:2017}. 

{Note that the above derivation of the equilibrium temperature at which $\lambda_\beta \approx a$ assumed that the characteristic wind speed $U$ was constant, which is not realistic over the broad range of equilibrium temperatures (as well as rotation periods, metallicities,  surface gravities, and frictional drag strengths) of hot Jupiters \citep{Roth:2024aa}. If I instead use the linear fit for equatorial jet speed with equilibrium temperature from \cite{Parmentier:2021tt} of $U_\mathrm{jet} = (6.2~T_\mathrm{eq}[K] - 5075)~\mathrm{m}~\mathrm{s}^{-1}$, I find a transition equilibrium temperature of $T_{\mathrm{eq},\beta} \approx 2166~\mathrm{K}$. The uncertainty in equilibrium temperature at which the Rhines scale is approximately equal to the planetary radius motivates exploring a broad range of equilibrium temperatures to assess whether dynamical changes can occur across this transition. However, in general} if $T_{eq} < T_{eq,\beta}$, then $\lambda_\beta > a$, and conversely if $T_{eq} > T_{eq,\beta}$ then $\lambda_\beta < a$. Given that hot Jupiters have $T_\mathrm{eq} \gtrsim 1000~\mathrm{K}$, the range of typical hot Jupiters are expected to cross this transition between small and large Rhines scales relative to the radius of the planet. 

One key caveat to this derivation is that the application of a Rhines scale to characterize the expected width of jets on of hot Jupiters assumes an inverse energy cascade \citep{Rhines:1975aa}. However, it is not clear a priori that hot Jupiters (or other tidally locked planets) undergo an inverse cascade in kinetic energy due to their large-scale day-to-night forcing causing the dominant component of the flow to be a wavenumber-1 planetary-scale wave pattern \citep{Showman_Polvani_2011,Hammond:2018aa}. As a result, the jet scale may be shaped by the forcing/damping leading to planetary-scale wave propagation and equatorward eddy momentum transport, rather than the $\beta$-effect \citep{Wang:2018}.
%(N.T. Lewis, personal communication). 
This would imply that the Rhines scale cannot be used to characterize the circulation regime of tidally locked planets. Instead the Rossby deformation radius may be a better indicator of circulation, given its direct relationship to planetary-scale gravity wave speeds.

\indent Note that in the above derivation the regime transition in which the Rhines scale becomes larger/smaller than the planetary radius is approximately equivalent to the regime transition where the Rossby number becomes larger/smaller than one \citep{Wang:2018}. If I take $\beta \approx f/a$, the Rhines scale is then $\lambda_\beta \approx \sqrt{\frac{Ua}{f}}$. One can then relate the Rhines scale and Rossby number as
\begin{equation}
    \label{eq:rolambdab}
    \frac{\lambda_\beta^2}{a^2} \approx \frac{U}{fa} \equiv \mathrm{Ro},
\end{equation}
assuming that the planetary radius is the relevant lengthscale of the flow. Thus, $\mathrm{Ro} = 1$ when $\lambda_\beta = a$, $\mathrm{Ro} > 1$ when $\lambda_\beta > a$, and $\mathrm{Ro} < 1$ when $\lambda_\beta < a$. As a result, even if the Rhines scale is not applicable to tidally locked exoplanets due to a lack of an inverse cascade \citep{Wang:2018}, the derivation here can also be used to determine the dynamical regime via the dominant planetary-scale momentum balance (i.e., advection balancing the day-to-night pressure gradient when $\mathrm{Ro} > 1$, and the Coriolis term balancing the day-night pressure gradient when $\mathrm{Ro} < 1$.)

\subsection{Rossby deformation radius}
%\cite{Haqq2018} define the transition between rapid and Rhines rotators to occur where 
Next, I explore the dynamical regime transition where the equatorial Rossby deformation radius $\lambda_R \sim a$ \citep{Haqq2018}, where 
\begin{equation}
    \lambda_R = \sqrt{\frac{NH}{2\beta}} \mathrm{,}
\end{equation}
and here $c_g \sim NH$ is the approximate peak equatorial Kelvin gravity wave speed \citep{showman_2013_doppler}. As above, I calculate the rotation period $P_R$ at this transition as a function of planetary radius, equilibrium temperature $T_\mathrm{eq}$, and the specific gas constant $R$, assuming an isothermal atmosphere at $T = T_\mathrm{eq}$\footnote{Note that for an isothermal atmosphere, $N = g/\sqrt{T c_p}$ and $H = RT/g$, implying that $\lambda_R = \sqrt{R\sqrt{T}/(2\sqrt{c_p}\beta)}$.}:
\begin{equation}
\begin{split}
    %P_R = \frac{4\pi a}{g \kappa} \mathrm{.} % old from jan 2024
    P_R \approx \frac{4\pi a \sqrt{c_p}}{R\sqrt{T_\mathrm{eq}}}~\mathrm{.}
\end{split}
\end{equation}
Here, $R$ and $c_p$ are the specific gas constant and specific heat capacity for a typical H$_2$ dominated hot Jupiter atmosphere.
%and $g$ is the planetary surface gravity. 

Equating $P_R = P_\mathrm{orb}$ from \Eq{eq:kepler}, I find the dependence of the transition equilibrium temperature between the regimes where the equatorial Rossby deformation radius is greater or larger than the planetary radius as a function of stellar, planetary, and atmospheric parameters:
\begin{equation}
\label{eq:teqr}
\begin{split}
    T_{eq,R} \approx & 796.1~\mathrm{K} \left(\frac{M_\star}{M_\odot}\right)^{-1/5} \left(\frac{L_\star}{L_\odot}\right)^{3/10} \\ & \times \left(\frac{a}{1.38~R_\mathrm{Jup,eq}}\right)^{-2/5}
    %old equation below from jan 2024
   % T_{eq,R} = & 115.4~\mathrm{K} \left(\frac{M_\star}{M_\odot}\right)^{-1/6} \left(\frac{L_\star}{L_\odot}\right)^{1/4} \\ & \left(\frac{g}{9.36~\mathrm{m}~\mathrm{s}^{-2}}\right)^{1/3} \left(\frac{a}{1.38~R_\mathrm{Jup,eq}}\right)^{-1/3}
    %old equation below before I fixed typo from Haqq-misra paper with gH not in another sqrt in Eq. 5
    %= & 3.739 \times 10^{4} ~\mathrm{K}  \left(\frac{M_\star}{M_\odot}\right)^{-1/4} \left(\frac{L_\star}{L_\odot}\right)^{3/8} \\ & \times \left(\frac{R_\mathrm{gas}}{3700~\mathrm{J}~\mathrm{kg}^{-1}~\mathrm{K}^{-1}}\right)^{1/2} \left(\frac{a}{R_\mathrm{{Jup,eq}}}\right)^{-1/2} 
     \mathrm{.}
\end{split}
\end{equation}
As before, all stellar values in \Eq{eq:teqr} are normalized to Solar and planetary parameters are normalized to HD 209458b-like values. If $T_{eq} < T_{eq,R}$, then $\lambda_R > a$, and conversely if $T_{eq} > T_{eq,R}$ then $\lambda_R < a$. 

Given that the equilibrium temperatures of hot Jupiters are $\gtrsim 1000~\mathrm{K}$, this implies that all typical hot Jupiters\footnote{Note that in this manuscript I am making a distinction between ``hot Jupiters'' and ``ultra-hot Jupiters.'' Ultra-hot Jupiters may have strong thermal inversions driven by absorption of incoming stellar radiation by metallic species \citep{Lothringer:2018aa} as well as larger specific gas constants due to molecular hydrogen dissociation \citep{Bell:2018aa}, both of which could increase $\lambda_R$. However, given that here I neglect the physics of hydrogen dissociation in this work for simplicity, I focus on the hot rather than ultra-hot regime throughout.} should have $\lambda_R < a$. {Note that unlike the Rhines scale, the Rossby deformation scale does not depend on the a priori unknown characteristic wind speed, and the theoretical uncertainty is linked instead to the thermal structure and atmospheric composition.} Additionally, even though warm Jupiters with $T_\mathrm{eq} \lesssim 796~\mathrm{K}$ would have $L_R > a$ if they were tidally locked, they are instead likely rapidly rotating due to the long tidal spin-down timescales at their larger orbital separations \citep{Guillot:1996}, leading to $L_R \ll a$ and a strongly banded circulation \citep{Showman:2014}.
%The transition between the regimes where the equatorial Rossby deformation radius is greater or less than the planetary radius occurs at equilibrium temperatures far below the regime of hot Jupiters. As a result, all hot Jupiters around Solar-type stars will have equatorial Rossby deformation radii that are smaller than the planetary radius. 
Note that as shown in \cite{Showman_2009}, the generalized Rossby deformation radius ($\lambda_D = c_g/f \sim NH/f$) is also smaller than the planetary radius for typical hot Jupiters. As a result, hot Jupiters are not expected to undergo a dynamical regime transition where the Rossby deformation radius $\lambda_R \gtrsim a$. 

\subsection{Dynamical regimes of tidally locked exoplanets}
\label{sec:dynregimes}
I now relate the transitions in the dynamical Rhines and equatorial Rossby deformation length scales more broadly to established dynamical regimes which have been published largely in the context of tidally locked temperate terrestrial exoplanets orbiting M dwarf stars \citep{Noda:2017aa,Haqq2018}. First, \cite{Haqq2018} define three dynamical regimes, in order of increasing relative rotation rate:
\begin{enumerate}
    \item Slow rotators: planets where both the Rhines and equatorial Rossby deformation length scales are larger than the planetary radius (i.e., $\lambda_\beta > a$, $\lambda_R > a$).
    \item Rhines rotators: planets where the Rhines scale is smaller than the planetary radius, but the Rossby deformation scale is  larger than the planetary radius (i.e., $\lambda_\beta < a$, $\lambda_R > a$). Note that this work, I will use the term intermediate rotator to refer to this regime, as hot Jupiters are not expected to lie in the Rhines rotator regime of \cite{Haqq2018}.
    \item Rapid rotators: planets where both the Rhines and equatorial Rossby deformation length scales are smaller than the planetary radius (i.e., $\lambda_\beta < a$, $\lambda_R < a$).
\end{enumerate}

As discussed above, I find that depending on their level of irradiation, host star type, planetary radius, and wind speeds, hot Jupiters can have Rhines scales that are less than or larger than the planetary radius. However, the equatorial Rossby deformation radii of hot Jupiters will be smaller than the planetary radius for reasonable ranges of host star type, atmospheric composition, 
%planetary gravity, 
and planetary radius, especially given that more irradiated hot Jupiters typically have larger radii \citep{Laughlin_2011,Thorngren:2018}. This implies that hot Jupiters can lie in either the rapid rotator regime or an analog of the Rhines rotator regime, which I term the ``Rossby'' rotator regime, where the Rossby deformation length is smaller than the planetary radius but the Rhines scale is larger than the planetary radius -- i.e., $\lambda_\beta > a$, $\lambda_R < a$. In the remainder of this work, I will classify both the Rhines and Rossby rotator regimes as ``intermediate'' rotators between the slow and rapid rotator limits. The primary reason why intermediate rotator hot Jupiters lie in the Rossby rather than Rhines rotator regime is because their equatorial jet speeds are typically several km~s$^{-1}$ rather than the hundreds of m~s$^{-1}$ level typical for temperate tidally locked terrestrial planets. 

Importantly, hot Jupiter dynamics does not extend into the slow rotator regime, unlike the dynamical regimes of temperate rocky planets orbiting late-type K dwarf stars or early-type M dwarf stars. The dynamics of hot Jupiters do not extend into the slow rotator regime because they have close-in and short period orbits (and as a result, hot equilibrium temperatures) and because they are found around earlier-type host stars than tidally locked temperate rocky planets. This has been demonstrated from previous large suites of tidally locked hot Jupiter GCM simulations, which predict superrotation regardless of the level of irradiation and rotation rate in the limit of weak frictional drag (e.g., \citealp{Tan:2019aa,Roth:2024aa}).

The four dynamical regimes (named Types I, II, III, IV) presented in \cite{Noda:2017aa} are based off of the dynamical characteristics from simulations of tidally locked terrestrial planets, rather than fundamental dynamical length scales. Type I is characterized by day-night flow and is equivalent to the slow rotator regime of \cite{Haqq2018}. Given that hot Jupiters do not lie in the slow rotator regime, they cannot be in the Type I regime of \cite{Noda:2017aa} except in cases where drag is sufficiently strong to damp the superrotating equatorial jet \citep{Perna_2010_1,Perez-Becker:2013fv,Beltz:2022aa}, causing the divergent (largely substellar to antistellar flow) component of the circulation to dominate over the rotational (consisting of the equatorial jet and waves) component \citep{Hammond:2021aa}. Instead, as shown in compilations of hot Jupiter GCMs \citep{Heng:2014b,Showman:2020rev} the typical circulation in drag-free simulations of hot Jupiters is equivalent to the Type II regime of \cite{Noda:2017aa}, with a superrotating equatorial jet and planetary-scale Rossby-Kelvin (Matsuno-Gill) wave pattern \citep{Pierrehumbert:2019vk}. 

Simulations of hot Jupiters covering a broad planetary and atmospheric parameter space (e.g., 
\citealp{Roth:2024aa}) do not typically exhibit circulation similar to Types III and IV of \cite{Noda:2017aa}, in which the superrotating equatorial jet becomes unstable (Type III) and transitions to two mid-latitude eastward jets (Type IV). As I will discuss later, the lack of Type III-IV circulation on hot Jupiters may be linked to the strong day-night forcing that drives their superrotating equatorial jet relative to the weaker forcing of temperate terrestrial planets. This is because the presence of superrotation has a strong dependence on day-night forcing strength \citep{Showman_Polvani_2011,Hammond:2020aa} and can undergo bifurcations with forcing \citep{Shell:2004}. Notably, the hysteresis found by \cite{sergeev:2022ab} at the boundary of the intermediate rotator and rapid rotator regimes corresponds to the transition between one superrotating equatorial jet (called ``Single Jet'' by \citealp{sergeev:2022ab}, equivalent to Type II of \citealp{Noda:2017aa}) and two mid-latitude eastward jets (called ``Double Jet'' by \citealp{sergeev:2022ab}, equivalent to Type IV of \citealp{Noda:2017aa}).

%old text below
%between Rhines and rapid rotators is irrelevant for hot Jupiters, as they all have equatorial Rossby deformation radii that are much larger than the planetary radius. This implies that regardless of forcing regime, hot Jupiters will not have a similar bifurcation in dynamical state as has been found in GCMs of TRAPPIST-1e \citep{sergeev:2022ab}. As a result, in this work we explore only the circulation of hot Jupiters at transition not explored in \cite{Haqq2018} where the Rhines scale can become smaller than the planetary radius while the Rossby deformation scale remains larger than the planetary radius.
%near the Rhines-slow rotator transition. 
%If $T_\mathrm{eq} > T_{eq,R}$, then the hot Jupiter is in the rapid rotator regime, and conversely if $T_\mathrm{eq} < T_{eq,R}$ the planet is in the Rhines rotator regime. 

\section{Numerical methods}
\label{sec:methods}
\subsection{Numerical setup}
 In this work, I focus on hot Jupiters where $T_{eq} \approx T_{eq,\beta}$ 
 %and $T_{eq} \approx T_{eq,Ro}$ 
 in order to study how the chosen initial conditions affect the resulting circulation regime at the transition between intermediate rotators (i.e., Rossby or Rhines rotators) and rapid rotators. I conduct GCM simulations with the cubed-sphere MITgcm \citep{Adcroft:2004},
 %varying initial temperature we use a setup that is an HD 209458b-like 
 %but with $a = R_\mathrm{Jup,eq}$, 
  using similar approaches to those outlined in detail in \cite{Komacek:2015} and \cite{Komacek:2017}. I do not include additional effects such as magnetic drag \citep{Perna_2010_1,Rogers:2020,Beltz:2022aa} or hydrogen dissociation and recombination \citep{Tan:2019aa,Roth:2021un}, but in some cases I do include a simplified uniform basal drag term \citep{Liu:2013}, the strength of which is freely varied. Specifically, I conduct two suites of GCM simulations, described further below:
 \begin{enumerate}
     \item Varying initial wind strength and direction for two basal drag strengths. This suite of GCMs builds directly upon \cite{Liu:2013}, extending their results to smaller basal drag timescales and longer model run times.  
     \item Varying initial temperature profile for a range of equilibrium temperatures, crossing from the intermediate rotator to rapid rotator regimes.
 \end{enumerate}
 \subsubsection{Varying initial wind profile}
 \label{sec:methodwind}
 In Suite 1, I study the effect of varying initial wind velocity using an idealized GCM setup adapted from \cite{Komacek:2015} with Newtonian cooling setting the heating/cooling rates in the dynamical core. I fix the planetary parameters to only study one planetary case, with an HD 209458b-like radiative equilibrium temperature profile \citep{Iro:2005}. In these Newtonian cooling models, the top boundary radiative equilibrium day-night temperature contrast and top boundary radiative timescales are free parameters, which I fix to $\Delta T_\mathrm{eq,top} = 1000~\mathrm{K}$ and $\tau_\mathrm{rad,top} = 10^4~\mathrm{s}$, respectively. The radiative equilibrium day-night temperature contrast then decreases logarithmically with increasing pressure until a maximum pressure of 10 bars, at presssures above which the radiative equilibrium day-night temperature contrast is fixed to zero. Similarly, the radiative timescale increases with increasing pressure as a power-law until a maximum pressure of 10 bars, at pressures above which it is fixed to $10^7~\mathrm{s}$. All other Newtonian cooling values and planetary parameters are kept the same as in \cite{Komacek:2015} (see Table \ref{tab:params}).

\begin{table}
%\vspace{0.5cm}
\setlength{\tabcolsep}{0pt}
\footnotesize
\begin{center}
%\resizebox{0.75\width}{!}{%
\begin{tabular}{ l l }
\hline
{\bf Parameter} & {\bf Value} \\
\hline
%\hline
%Visible opacity & $10^{0.0478(\mathrm{log}_{10}p[\mathrm{Pa}])^2 - 0.1366(\mathrm{log}_{10}p[\mathrm{Pa}]) - 3.2095}$ & m$^2$~kg$^{-1}$ \\
%Infrared opacity & $\mathrm{max}\left(10^{0.0498(\mathrm{log}_{10}p[\mathrm{Pa}])^2 - 0.1329(\mathrm{log}_{10}p[\mathrm{Pa}]) - 2.9457}, 10^{-3}\right)$ & m$^2$~kg$^{-1}$ \\ 
\hline 
Suite 1 parameters \\
\hline 
Initial peak zonal wind velocity & [0, +1, -1] m$~\mathrm{s}^{-1}$ \\  
Basal drag constant ($k_F$) & [$10^{-2}$, $10^{-3}$]~$\mathrm{day}^{-1}$ \\
Rotation period & $3.5~\mathrm{days}$\\
Radius & $9.43 \times 10^{7}~\mathrm{m}$ \\
%\hline
Gravity & 9.36 m~s$^{-2}$\\
%\hline
%\hline
%\hline
Top radiative equilibrium  & $1000~\mathrm{K}$ \\
\hspace{0.04cm} day-night temperature contrast & \\ 
Maximum pressure of & $10^{-3}~\mathrm{bars}$ \\ 
\hspace{0.04cm} top day-night temperature contrast & \\
Bottom radiative equilibrium  & $0~\mathrm{K}$ \\
\hspace{0.04cm} day-night temperature contrast & \\ 
Minimum pressure of bottom & $10 ~\mathrm{bars}$ \\ 
\hspace{0.04cm} day-night temperature contrast & \\
Top radiative timescale & $10^4~\mathrm{s}$ \\
Maximum pressure of & $10^{-2} ~\mathrm{bars}$ \\ 
\hspace{0.04cm} top radiative timescale & \\
Bottom radiative timescale & $10^7~\mathrm{s}$ \\
Minimum pressure of & $10~\mathrm{bars}$ \\
\hspace{0.04cm} bottom radiative timescale & \\ 
Basal drag minimum pressure & $10~\mathrm{bars}$ \\
Horizontal resolution & C16 \\
%\hline 
%\hline 
%\hline 
%\hline
Newtonian cooling time step & 45 s\\
\hline 
Suite 2 parameters \\ 
\hline 
\textit{Suite 2(a)} \\
Equilibrium temperature & $[1100, 1200, 1300, 1400, 1500]~\mathrm{K}$ \\
Rotation period &  [$5.93, 4.57, 3.59, 2.88, 2.34]~\mathrm{days}$ \\ 
Dynamical time step & 15 s \\
Radiative time step & 150 s\\

\textit{Suite 2(b)} \\
Equilibrium temperature & $[1600, 1700, 1800, 1900, 2000,$ \\ 
& $ 2100, 2200, 2300, 2400, 2500]~\mathrm{K}$ \\
Rotation period &  $[1.93, 1.61, 1.35, 1.15, 0.99,$ \\
& $ 0.85, 0.74, 0.65, 0.57, 0.51]~\mathrm{days}$ \\ 
Dynamical time step & 5-15 s \\
Radiative time step & 20-30 s\\
\textit{All of Suite 2} \\
Initial temperature & [$T_\mathrm{eq}/\sqrt{2}$, $\sqrt{2} T_\mathrm{eq}$] \\ 
Radius & $9.87 \times 10^7~\mathrm{m}$  \\
%\hline
Gravity & 9.30 m~s$^{-2}$\\
%\hline
Drag constant ($k_F$) & 0 \\
Internal temperature & $100~\mathrm{K}$ \\ 
%\hline
Horizontal resolution & C32 \\
%\hline 
%\hline
\hline
Parameters common to all simulations \\
\hline
Specific heat capacity & $1.3 \times 10^4$ J~kg$^{-1}$~K$^{-1}$ \\
%\hline
Specific gas constant & 3714 J~kg$^{-1}$~K$^{-1}$ \\
%\hline
Vertical layers & 40 \\
%\hline 
%\hline 
Upper boundary pressure & 0.183 mbars \\
Lower boundary pressure & 200 bars \\
%\hline 
Shapiro filter order & 4 \\
Shapiro filter timescale & 25 s \\ 

%Vertical Velocity ($\omega$) & $\omega < 0~\mathrm{Pa}~\mathrm{s}^{-1}$ \\
%\hline
%&  Hurricane: \\
%& \\
%Maximum Potential Intensity ($u_p$) & Tropical Storm: \\
%& $33 > u_p \ge 18~\mathrm{m}~\mathrm{s}^{-1}$ \\
%& \\
%& Tropical Depression: \\
%& $u_p < 18~\mathrm{m}~\mathrm{s}^{-1}$\\
\hline
\end{tabular}%
%}
\caption{Planetary and atmospheric properties assumed for both suites of GCM simulations presented in this work, along with atmospheric and numerical parameter choices common to all simulations.}
\label{tab:params}
\end{center}
\end{table}

\begin{figure*}
    \centering
     \includegraphics[width=1\textwidth]{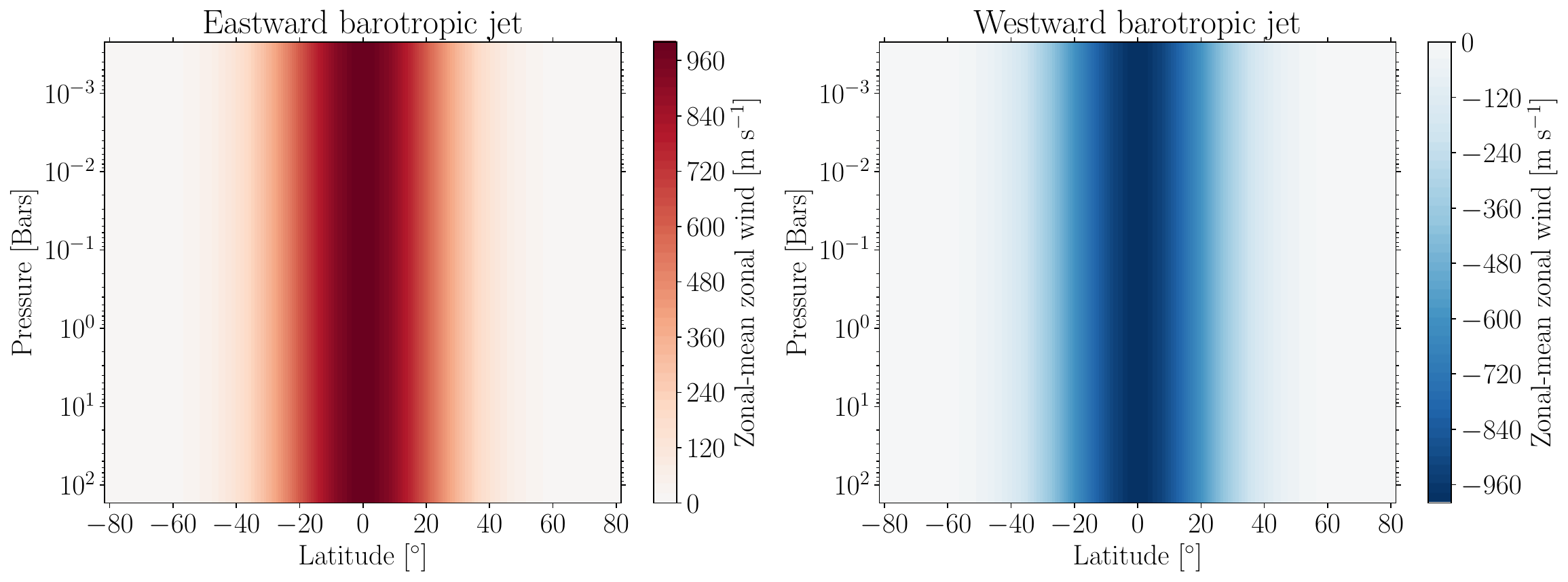}
     \caption{Initial zonal-mean wind profiles for GCM cases in Suite 1 with varying initial wind direction. The two maps share a color scale, with red representing eastward zonal-mean zonal wind speeds, blue representing westward wind speeds, and white zero zonal-mean zonal wind. In both cases, the initial jet is prescribed to be barotropic, with no variation in pressure.}
     \label{fig:initjet}
 \end{figure*}

 Similarly to the shallow-water and 3D GCM models with Newtonian cooling of \cite{Liu:2013}, I conduct simulations for three separate assumptions of initial wind speeds, either no winds (``rest''), an everywhere eastward barotropic flow, or a westward barotopic flow. In both of the latter cases, the flow is initialized with a maximum velocity at the equator of $+1~\mathrm{km}~\mathrm{s}^{-1}$ (``eastward'') and $-1~\mathrm{km}~\mathrm{s}^{-1}$ (``westward''), and the winds decay with latitude. Figure \ref{fig:initjet} shows the initial wind profiles for these two cases with eastward and westward initial flow. In both cases, the wind profiles vary only as a function of latitude and are fixed in both longitude and pressure. 

In this set of simulations with varying initial wind profiles, I also include a basal drag which is horizontally uniform but increases in strength from a given pressure toward the bottom of the domain. I consider two levels of basal drag, with a maximum strength a factor of ten and a hundred times weaker than the nominal value in \cite{Liu:2013}, as described further below. I do so in order to determine the level of impact of drag on the atmospheric circulation, extending the results of \cite{Liu:2013} to include a full suite of simulations with weaker frictional drag as well as varying the initial wind profile. Note that \cite{Liu:2013} do state that they explore cases with varying frictional drag strength, but they do not present a systematic exploration of the impact of frictional drag strength on the resulting dynamics. 

The ``basal'' drag is parameterized as in \cite{Liu:2013}. This drag is only applied in the deep atmosphere at pressures above 10 bar, motivated by the frictional interaction between the atmosphere and the relatively quiescent interior. The drag force is parameterized as a Rayleigh drag
 \begin{equation}
     \mathcal{F}_d = -k_F {\bf v} = -\frac{\bf v}{\tau_\mathrm{drag}},
 \end{equation}
 where $k_F$ increases linearly with pressure from zero at 10 bars to a maximum value at the bottom of the domain at a pressure of 200 bars \citep{Komacek:2015}. I conduct simulations with two values of the maximum $k_F$ (i.e., the drag constant at 200 bars) of $10^{-2}~\mathrm{day}^{-1}$ and $ 10^{-3}~\mathrm{day}^{-1}$. These values of $k_F$ correspond to equivalent minimum basal drag timescales of $\tau_\mathrm{drag} = 8.64 \times 10^{6}~\mathrm{s}$ and $ 8.64 \times 10^{7}~\mathrm{s}$. These are weaker basal drag coefficients (longer drag timescales) than used in \cite{Liu:2013}, who take $k_F = 10^{-1}~\mathrm{day}^{-1}$ for their primary suite of 3D models. This results in a total of six simulations in Suite 1, three cases with varying initial winds for each of the two chosen drag strengths. 

\subsubsection{Varying initial temperature}
\label{sec:methodstemp}
  In Suite 2, I conduct a suite of simulations crossing the intermediate rotator to rapid rotator transition with varying irradiation, here written as the full-redistribution equilibrium temperature $T_\mathrm{eq}$. For each case at a given $T_\mathrm{eq}$, I conduct two simulations with varying initial temperature, one ``hot start'' and one ``cold start,''\footnote{As these are not evolutionary calculations, the nomenclature is similar but not equivalent to the hot start and cold start in gas giant formation models (e.g., \citealp{Berardo:2017aa,Youdin:2025aa}).} in order to determine whether hot Jupiters could exhibit a dependence on initial conditions at the intermediate-rapid rotator regime transition as was found for TRAPPIST-1e in \cite{sergeev:2022ab}.
  %(see details in \Sec{sec:sweep}). 
  In all simulations for Suite 2, I couple the MITgcm to the double-gray version of the DISORT \citep{Kylling:1992,Stamnes:2027} two-stream radiative transfer scheme, as used in \cite{Komacek:2017}. The model setup is similar to that in \cite{Komacek:2017}, including the choices of double-gray visible and infrared band opacities. This setup is relevant for an HD 209458b-like H$_2$/He dominated atmosphere, except for the choices of equilibrium temperature and initial temperature profile discussed further below (also see Table \ref{tab:params}). I do not include frictional drag at any point in the atmosphere for these cases in Suite 2 with a varying initial temperature profile, because drag (especially deep basal drag) has previously been shown to potentially limit hysteresis in hot Jupiter GCM simulations (\citealp{Liu:2013}, see also Suite 1).
  
  I conduct two sub-grids of simulations in Suite 2. In Suite 2(a), I conduct simulations varying instellation to cross $T_{eq,\beta}$ in steps of 100 K (1100 K, 1200 K, 1300 K, 1400 K, 1500 K crossing the expected $T_{eq,\beta} \approx 1300~\mathrm{K}$ assuming that wind speed does not scale steeply with instellation in this regime). I keep the simulation parameter space focused on relatively cool temperatures in order to avoid the impacts of hydrogen dissociation and recombination \citep{Bell:2018aa,Tan:2019aa,Roth:2021un} and Lorentz forces due to the internal magnetic field coupling to the flow \citep{Perna_2010_1,batygin_2013,Rauscher_2013,Rogers:2020,Beltz:2022aa} at higher temperatures, which are not included in these idealized simulations. In these simulations, I keep the orbital period equal to the rotation period and include the variation in rotation period consistently with Kepler's third law assuming a Sun-like star -- as a result, the varying $T_\mathrm{eq}$ corresponds to varying $P_\mathrm{orb} = P_\mathrm{rot}$ ($5.93~\mathrm{days}, 4.57~\mathrm{days}, 3.59~\mathrm{days}, 2.88~\mathrm{days}, 2.34~\mathrm{days}$, decreasing with increasing $T_\mathrm{eq}$).
  %different initial conditions. 
 For each set of ($T_\mathrm{eq}, P_\mathrm{rot}$) I vary the isothermal initial temperature profile, where cold start cases have $T = T_\mathrm{eq}/\sqrt{2}$ and hot start cases have $T = T_\mathrm{irr} = \sqrt{2}T_\mathrm{eq}$. This results in a total of 10 simulations, with two initial temperature profiles for each of the five assumed combinations of ($T_\mathrm{eq}, P_\mathrm{rot}$). 

 In Suite 2(b), I extend the grid of simulations in Suite 2(a) to cover the potential for a higher transition equilibrium temperature if the linear fit for the equatorial jet speed from \cite{Parmentier:2021tt} is used rather than a fixed jet speed as in \Eq{eq:teqbeta}. This extension covers equilibrium temperatures from 1600-2500 K again in 100 K increments, with both hot and cold start cases using the same prescription as in Suite 2(a), resulting in 20 additional simulations for a total of 30 in Suite 2. These simulations in Suite 2(b) also include a varying planetary rotation period with equilibrium temperature assuming a Sun-like star as in Suite 2(a). Table \ref{tab:params} details the planetary and numerical parameter choices for both Suites 2(a) and 2(b).
  %probably need to add a summary table showing Suite 1, Suite 2 assumptions, done.
% \subsection{Parameter sweep}
 %\label{sec:sweep}
 %make table stating grid #, parameters (initial and forcing/damping) varied, values chosen, done.

 \subsection{Numerical details}
 All simulations presented in the manuscript are conducted to sufficient run time to enable convergence in wind profiles, as measured via the domain-integrated and level-by-level kinetic energy. Simulations in Suite 1 with varying initial wind speed are run to $\sim 25,000$ Earth days, with these long run timescales enabled by the idealized Newtonian cooling scheme. Note that these model runtimes are significantly longer than those presented in \cite{Liu:2013}, which were only conducted to $\sim 1,000$ Earth days. The simulations in Suite 2(a) with varying initial temperature profiles are each conducted for $\sim 5,000$ days to reach convergence, while the simulations in Suite 2(b) are each conducted to a shorter end model time of $\sim 1,500$ days. 
 
 The simulations with varying initial wind speed have a low cubed-sphere resolution of C16 (approximately 64 x 32 x 40 in Cartesian coordinates). Simulations with varying initial temperature have a higher resolution of C32 (approximately 128 x 64 x 40). All simulations in both model Suites 1 and 2(a) have a dynamical timestep of 15 seconds, while some simulations in Suite 2(b) have shorter dynamical time steps as low as 5 seconds as well as shorter radiative time steps as low as 20 seconds for numerical stability. All numerical results shown in the following sections for cases in both Suite 1 and Suite 2(a) are averaged over the last 500 days of model time, while those in Suite 2(b) are averaged over the last 350 days of model time. 

 %Additionally can vary initial wind field (no flow vs. westward vs. eastward jet - 10 more simulations if warranted). Can make temperature grid finer if we find bifurcation. 

 \section{Numerical results}
 \label{sec:results}
 \subsection{Varying initial wind profile}
\begin{figure*}
    \centering
     \includegraphics[width=0.95\textwidth]{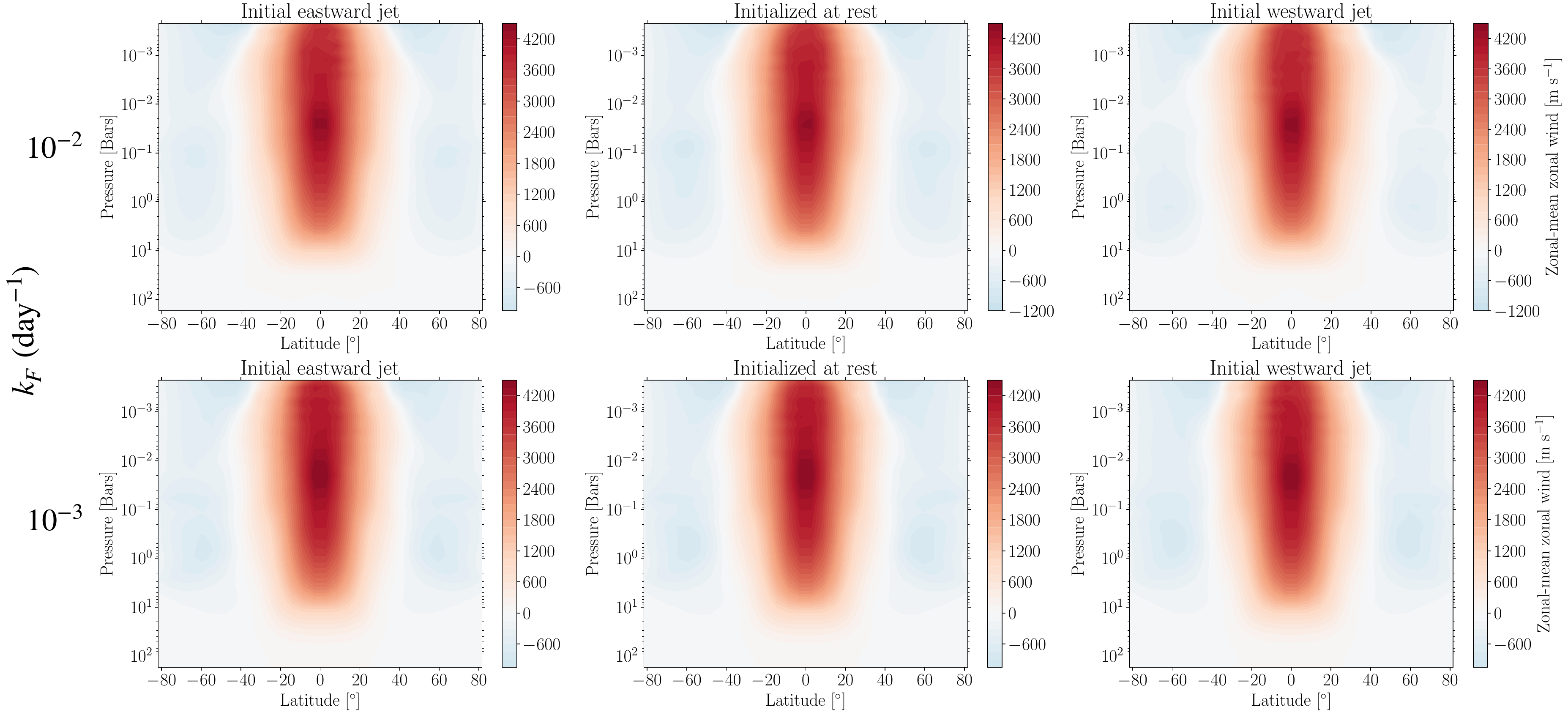}
     \caption{Zonal-mean zonal wind profiles from simulations with different initial wind patterns. The left-hand column shows simulations initialized with a barotropic eastward jet, the middle column shows simulations initialized at rest, and the right column shows simulations initialized with a barotropic westward jet. The top row corresponds to simulations with a maximum basal drag parameter of $k_F = 10^{-2}~\mathrm{day}^{-1}$, and the bottom   row corresponds to simulations with a weaker basal drag parameter of $k_F = 10^{-3}~\mathrm{day}^{-1}$. All plots share colormap limits. I find that the zonal-mean zonal wind from simulations for a given drag parameter with different initial wind profiles are nearly identical.}
     \label{fig:varyjet_zonalwind}
 \end{figure*}

 I first describe results from the suite of simulations with varying initial wind profile for two separate basal drag strengths, the model setup of which is outlined in detail in \Sec{sec:methodwind}. \Fig{fig:varyjet_zonalwind} shows zonal-mean zonal wind profiles from each of the six simulations with varying initial wind profile (columns) and maximum basal frictional drag coefficient (rows). 
All simulations display the characteristic hot Jupiter zonal-mean circulation pattern characterized by a superrotating equatorial jet \citep{Showman_Polvani_2011}, regardless of whether the initial wind profile is superrotating or subrotating.

The strength of the simulated jet depends slightly on the maximum drag coefficient $k_F$, as simulations with stronger drag (higher $k_F$) have weaker peak wind speeds. For instance, the case begun from rest with $k_F = 10^{-2}~\mathrm{day}^{-1}$ has a peak zonal-mean zonal wind speed of $4366.9~\mathrm{m}~\mathrm{s}^{-1}$, while the case started from rest with $k_F = 10^{-3}~\mathrm{day}^{-1}$ has a faster peak zonal-mean zonal wind speed of $4495.1\mathrm{m}~\mathrm{s}^{-1}$ (a $2.85\%$ increase). However, all simulations at a given drag strength show zonal-mean zonal wind profiles which are nearly identical regardless of initial wind profiles. The largest difference in maximum zonal-mean zonal wind speed between simulations at a given drag strength is $32.4~\mathrm{m}~\mathrm{s}^{-1}$ (a $0.742\%$ variation), which occurs in the cases with $k_F = 10^{-2}~\mathrm{day}^{-1}$. Meanwhile, the cases with $k_F = 10^{-3}~\mathrm{day}^{-1}$ show at most a $0.216\%$ variation in zonal-mean zonal winds. 

\begin{figure*}
    \centering
     \includegraphics[width=0.95\textwidth]{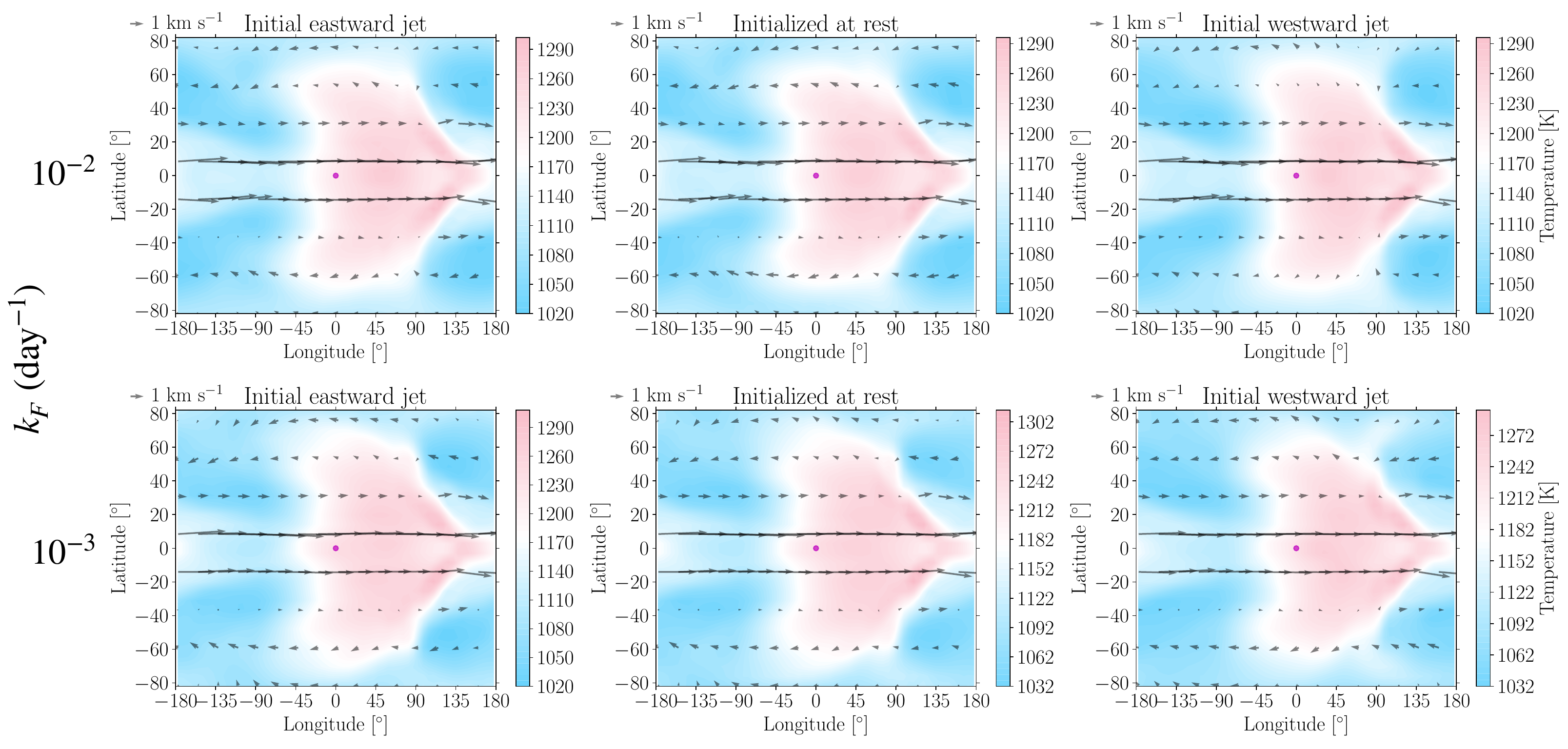}
     \caption{Temperature (colors) and wind (quivers) maps at 81.5 mbar pressure from simulations with different initial wind patterns. The left-hand column shows simulations initialized with a barotropic eastward jet, the middle column simulations initialized at rest, and the right column simulations initialized with a barotropic westward jet. The top row corresponds to simulations with maximum basal drag parameter of $k_F = 10^{-2}~\mathrm{day}^{-1}$, and the bottom row corresponds to simulations with a weaker maximum drag parameter of $k_F = 10^{-3}~\mathrm{day}^{-1}$. All plots share colormap limits, but the colorbar limits are kept separate for each plot in order to display small inter-model differences. I find that the temperature pattern from simulations with different initial wind profiles are nearly identical for a given drag parameter, as expected given their similar jet structure.}
     \label{fig:varyjet_tempwind}
 \end{figure*}

For completeness, I also show the near-photospheric (81.5 mbar) temperature and wind maps from the same six simulations with varying initial wind profile and frictional drag strength in \Fig{fig:varyjet_tempwind}.
As for the zonal-mean zonal wind speeds, the wind and temperature patterns on an isobar display very small inter-model differences. All cases have an eastward hot spot offset driven by the superrotating equatorial jet Doppler-shifting the planetary-scale wave pattern \citep{Hammond:2018aa}. All cases also have their coldest regions at Rossby gyres in the mid-latitudes, and a nearly identical flow pattern everywhere for a given $k_F$. Overall, these small inter-model differences in jet speeds and temperature structures with varying initial wind profiles agree well with the findings of \cite{Liu:2013}, extending their results to weaker drag and longer run-times. This implies that the strength of basal drag does not significantly impact the near-photospheric jet structure of hot Jupiters. Overall, I find no evidence from this limited set of GCM simulations that the atmospheric circulation in 3D simulations of hot Jupiter dynamics displays a dependence on the initial wind profile.

 \subsection{Varying initial temperature profile}
 \subsubsection{Time-averaged temperature and winds}
\begin{figure}
    \centering
    \includegraphics[width=0.5\textwidth]{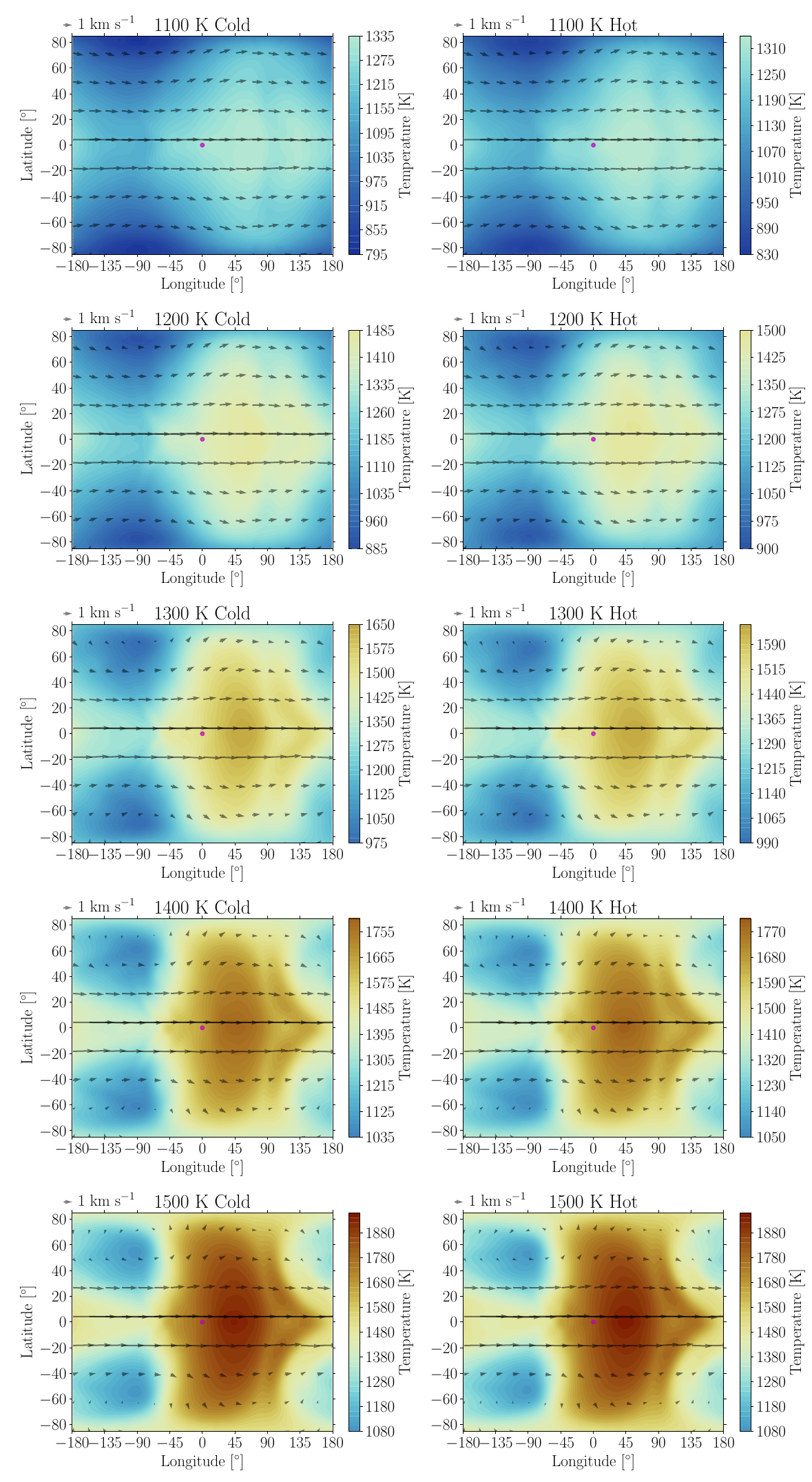}
    \caption{Temperature (colors) and wind (quivers) maps at 81.5 mbar pressure from Suite 2(a) with varying equilibrium temperature (rows) and initial conditions (columns). Each map shares a color scale with \Fig{fig:tempwind_2b}, and the colorbar beside the map displays the range of temperatures contained within the map. A $1~\mathrm{km}~\mathrm{s}^{-1}$ quiver is displayed to the upper-left of each map. I find that the near-photospheric temperature and winds are nearly identical in both the cases with cold and hot initial conditions. This implies that there is no hysteresis in the photospheric climate state across the intermediate rotator-rapid rotator transition using the dynamical regime definitions of \cite{Haqq2018}. }
    \label{fig:tempwind}
\end{figure}

\begin{figure*}
    \centering
    \includegraphics[width=0.95\textwidth]{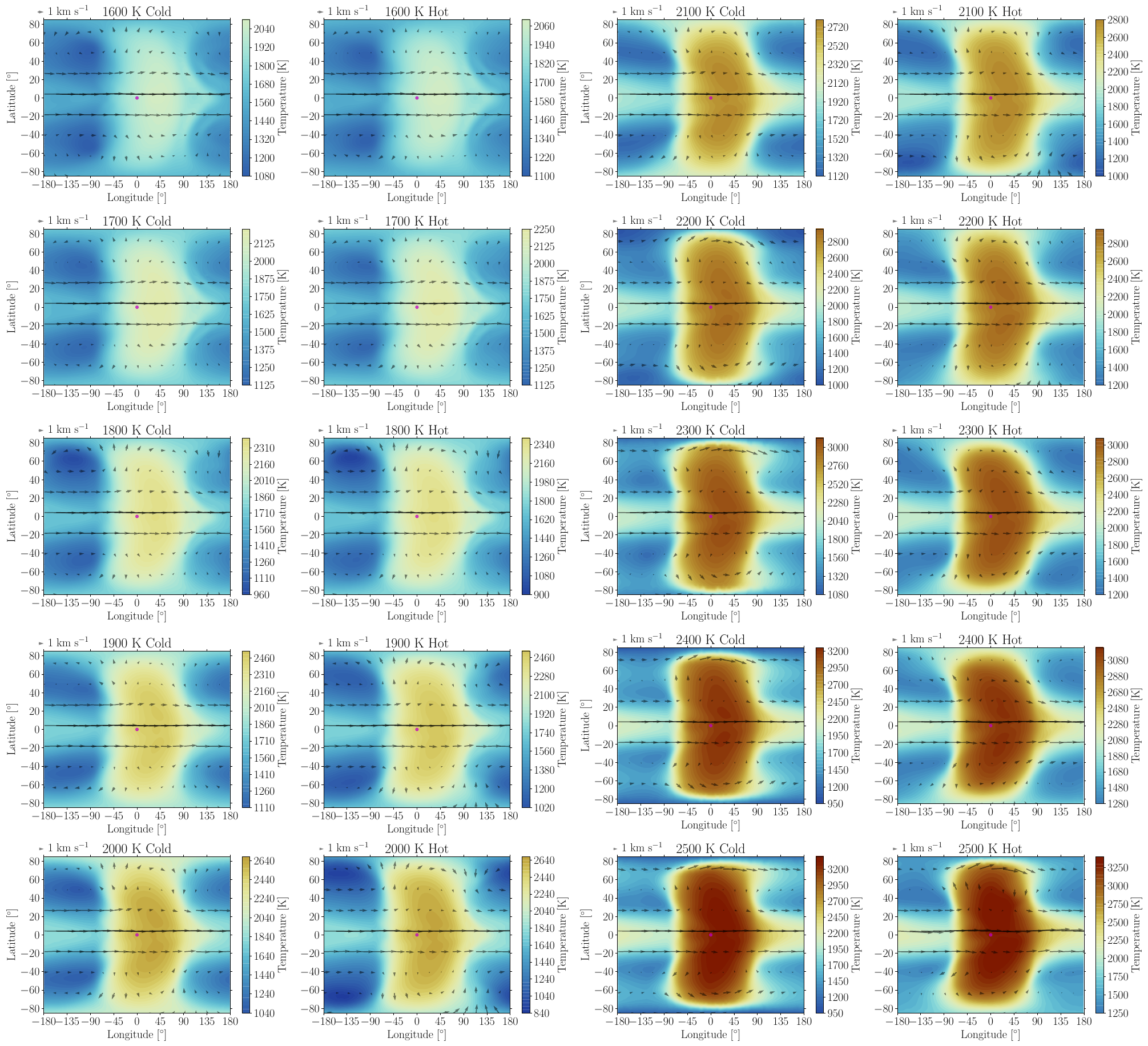}
    \caption{As in Figure 
    \ref{fig:tempwind}, but here showing temperature (colors) and wind (quivers) maps at 81.5 mbar pressure from the simulations in Suite 2(b) with varying equilibrium temperature (rows) and initial conditions (columns). Each map shares a color scale with \Fig{fig:tempwind}, and the colorbar beside the map displays the range of temperatures contained within the map. A $1~\mathrm{km}~\mathrm{s}^{-1}$ quiver is displayed to the upper-left of each map.}
    \label{fig:tempwind_2b}
\end{figure*}

%The initial grid of planets varying $T_\mathrm{eq}$ from 1100-1500 K for cold and hot starts was conducted to $\sim 5,000~\mathrm{days}$. 
I next turn to describe results from the second suite of simulations with varying initial temperature profile, the model setup of which is outlined in \Sec{sec:methodstemp}. \Fig{fig:tempwind} shows temperature and wind maps near the photosphere from all ten GCM simulations in Suite 2(a) with varying equilibrium temperature (rows) from $1100-1500~\mathrm{K}$ for cold and hot initial temperature profiles (columns), and \Fig{fig:tempwind_2b} shows the equivalent temperature and wind maps from the twenty GCM simulations in Suite 2(b).  

Increasing equilibrium temperature increases the dayside temperature and day-to-night temperature contrast leading to a corresponding increase in the speed of atmospheric winds, as expected \citep{Perez-Becker:2013fv}. Simulations for both cold and hot start cases at a given equilibrium temperature (i.e., in a given row) display similar horizontal temperature structures and wind patterns. This implies that there is no key change in the near-photospheric thermal structure across the theoretical regime transition where the Rhines scale is greater than or less than the planetary radius previously derived in \Eq{eq:teqbeta} (Suite 2a), or when including a linear dependence of wind speed on equilibrium temperature (Suite 2b).

\begin{figure}
    \centering
    \includegraphics[width=0.5\textwidth]{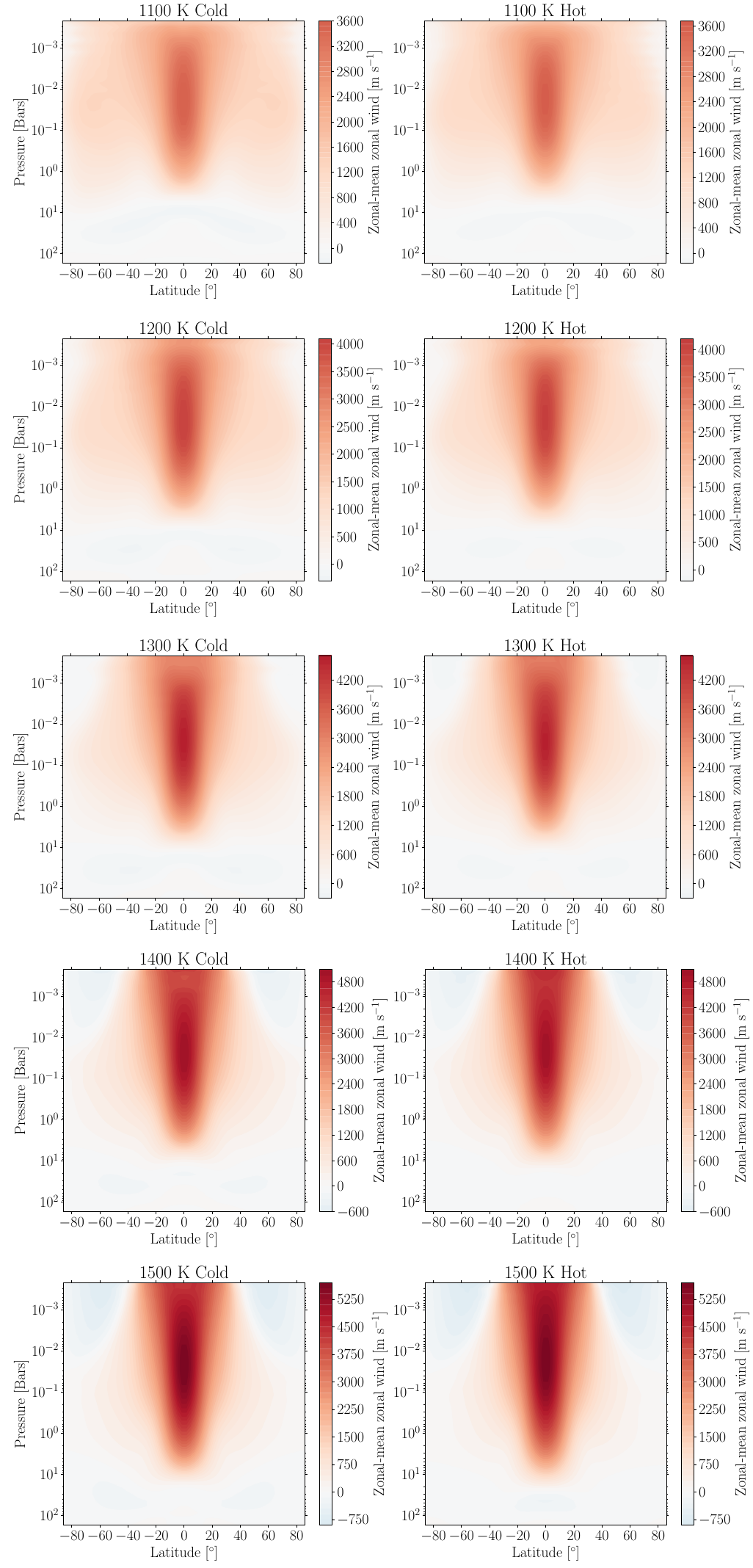}
    \caption{Zonal-mean zonal wind from Suite 2(a) with varying equilibrium temperature (rows) and initial conditions (columns). Each map shares a color scale with \Fig{fig:zonalwind_2b}, and the colorbar displays the range of zonal-mean zonal wind speeds in each plot. White corresponds to zero zonal-mean zonal wind. I find that all cases have a superrotating eastward equatorial jet, with the zonal-mean zonal wind pattern nearly identical between the cases with cold and hot initial conditions.}
    \label{fig:zonalwind}
\end{figure}

\begin{figure*}
    \centering
    \includegraphics[width=0.95\textwidth]{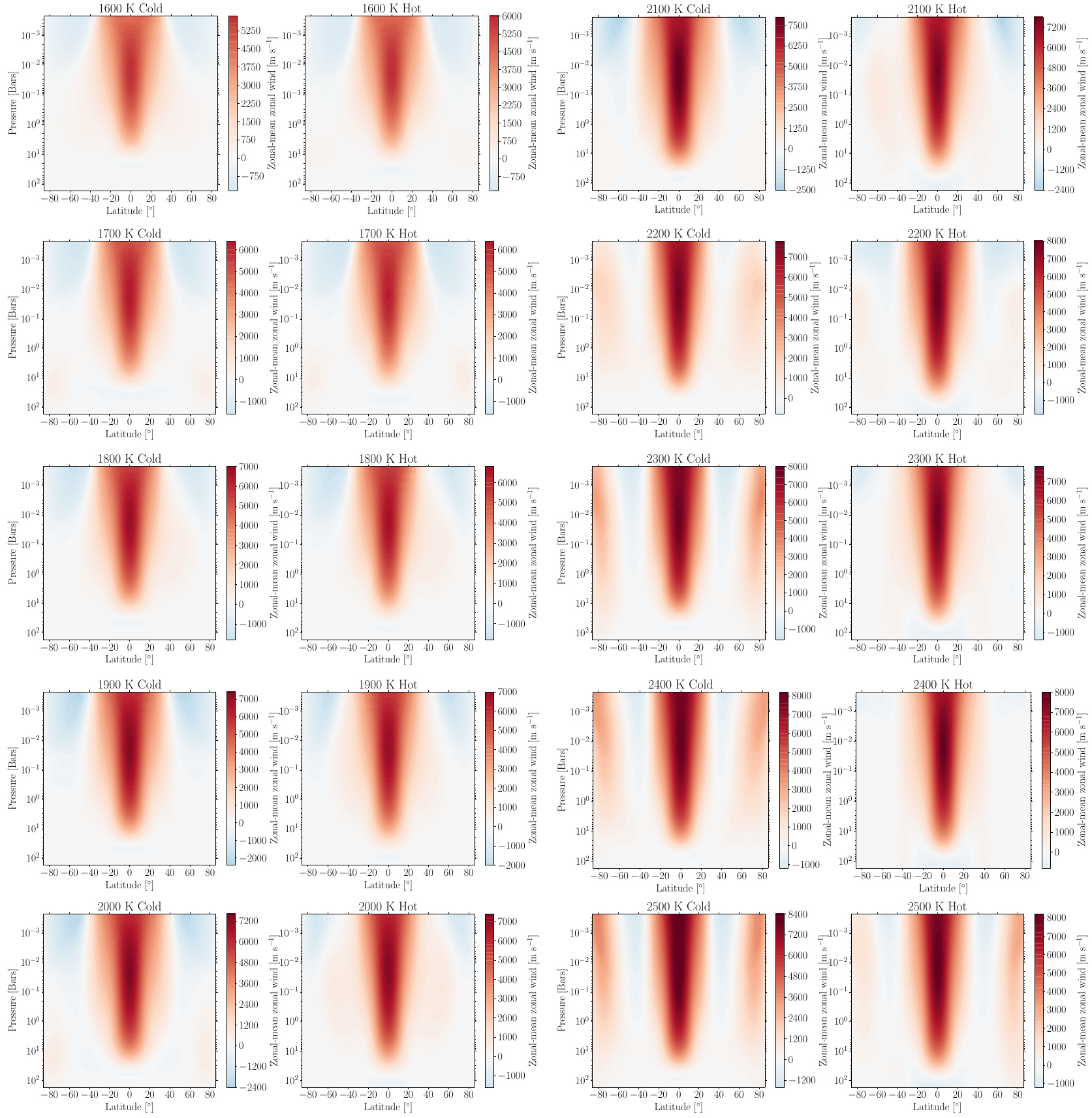}
    \caption{As in Figure \ref{fig:zonalwind}, but here showing zonal-mean zonal wind from Suite 2(b) with varying equilibrium temperature (rows) and initial conditions (columns). Each map shares a color scale with \Fig{fig:zonalwind}, and the colorbar displays the range of zonal-mean zonal wind speeds in each plot. White corresponds to zero zonal-mean zonal wind.}
    \label{fig:zonalwind_2b}
\end{figure*}

The corresponding zonal-mean zonal winds as a function of latitude and pressure from simulations with varying equilibrium temperature for cold and hot initial thermal profiles are shown for Suite 2(a) in \Fig{fig:zonalwind} and for Suite 2(b) in \Fig{fig:zonalwind_2b}.
The zonal-mean zonal wind increases with increasing equilibrium temperature due to the increasing day-to-night forcing amplitude, as expected. Additionally, there is no significant change in the zonal-mean zonal wind structure between cold and hot start cases, as was found for the near-photospheric horizontal thermal structure. This implies that there is no bifurcation between single jet and double jet regimes on hot Jupiters crossing the theoretical $L_\beta \sim a$ threshold. %Because $L_R < a$ in all cases, 
Overall, I find no significant evidence for a dependence of the jet structure on initial conditions for hot Jupiters that was found in the terrestrial planet simulations of \cite{sergeev:2022ab}. 
However, note that this may not apply to \textit{ultra-hot} Jupiters, as the high-latitude flow (but not the equatorial jet) may be sensitive to initial conditions and including molecular hydrogen dissociation and recombination, which is neglected here, may affect the potential for hysteresis (see Section \ref{sec:limitations}).

\subsubsection{Spatiotemporal averages}
\begin{figure}
\centering
    \includegraphics[width=0.475\textwidth]{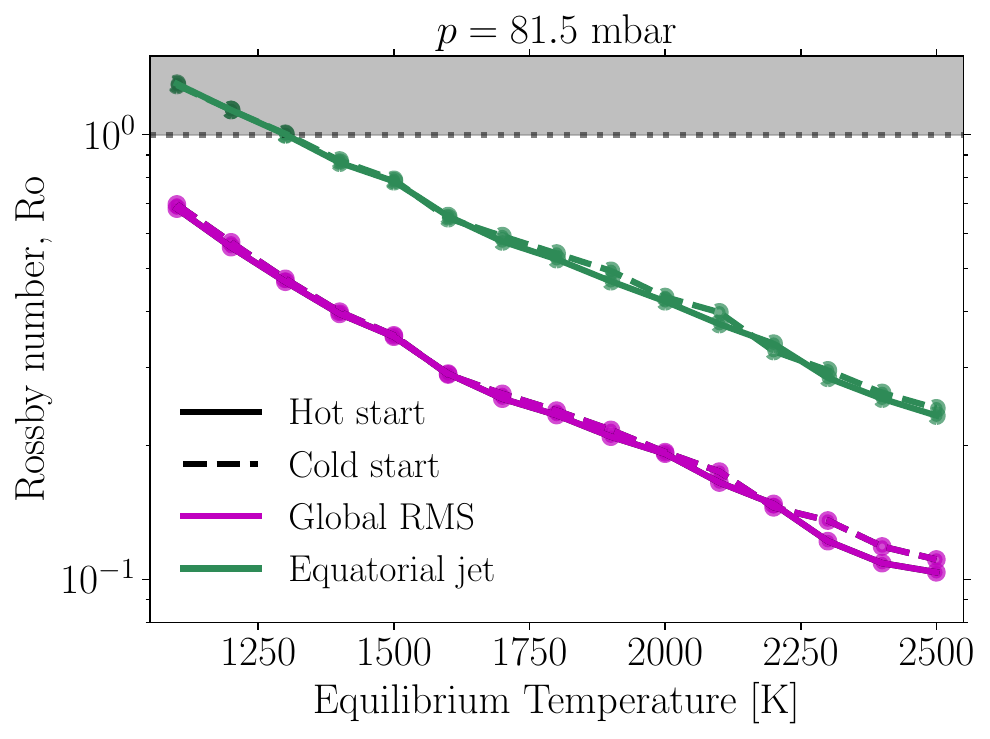}
    \caption{Rossby number 
    %and Rhines scale normalized to the planetary radius (bottom) 
    numerically calculated from the GCM simulations in Suite 2. Solid lines connecting dots correspond to the simulations initialized with a hot start while dashed lines correspond to the cold start cases. The green curves correspond to numerical estimates of the Rossby number 
    %and Rhines scale 
    using the zonal-mean zonal wind speed averaged within $10^\circ$ of the equator, while the magneta curves correspond to numerical estimates using the global root-mean-square (RMS) wind speed. The dotted horizontal line 
    %on each plot 
    displays $\mathrm{Ro} = 1$, which as shown in \Eq{eq:rolambdab} is approximately equivalent to the point where $\lambda_\beta/a = 1$ given that $\mathrm{Ro} \approx (\lambda_\beta/a)^2$. The gray shaded region above the $\mathrm{Ro} = 1$ line corresponds to the regime where the Rossby number and as a result the} Rhines scale are large (i.e., the intermediate rotator regime described in Section \ref{sec:dynregimes}). I find that when using the equatorial jet speed as the characteristic wind scale, GCM simulations of hot Jupiters can lie in the intermediate rotation regime for $T_\mathrm{eq} \lesssim 1300~\mathrm{K}$, as expected from the scaling theory presented in Section \ref{sec:dynregime}.
    \label{fig:RoRh}
\end{figure}

The above description of results from Suites 2(a) and 2(b) demonstrates that there is no dependence of the temperature or wind pattern on the initial thermal structure from simulations covering equilibrium temperatures from $1100-2500~\mathrm{K}$. However, these simulations have not been demonstrated to cross a predicted regime transition from intermediate to rapid rotators with varying equilibrium temperature. To demonstrate that these simulations cross a theoretical regime transition, in \Fig{fig:RoRh} I show the numerically calculated Rossby number 
%and Rhines scale 
using the GCM simulations from Suite 2. Note that $\mathrm{Ro} \approx (\lambda_\beta/a)^2$ from \Eq{eq:rolambdab}. As a result, the shaded region corresponds to the ``intermediate'' rotator regime with $\mathrm{Ro} >1$ and equivalently $\lambda_\beta/a > 1$, while values of $\mathrm{Ro} < 1$ and equivalently $\lambda_\beta/a < 1$ correspond to the rapid rotator regime. I find that as expected from the scaling in \Eq{eq:teqbeta}, the transition between the intermediate and rapid rotator regimes occurs at $T_\mathrm{eq} \approx 1300~\mathrm{K}$ if the equatorial jet speed is taken to be the characteristic wind speed. Conversely, the slower global RMS wind speeds imply that if the global RMS winds are taken as the characteristic wind speeds then all cases lie in the rapid rotator regime with $\mathrm{Ro} < 1$ and $\lambda_\beta/a < 1$. However, note that the fact that the simulations cross the $\mathrm{Ro} = 1$ and equivalently $\lambda_\beta/a \approx 1$ threshold does not necessarily imply that they cross an actual dynamical regime transition as in \cite{Haqq2018,Wang:2018}, and \cite{sergeev:2022ab}, as all simulations have broadly the same dynamical characteristics including a single superrotating equatorial jet.

\begin{figure}
    \centering
    \includegraphics[width=0.5\textwidth]{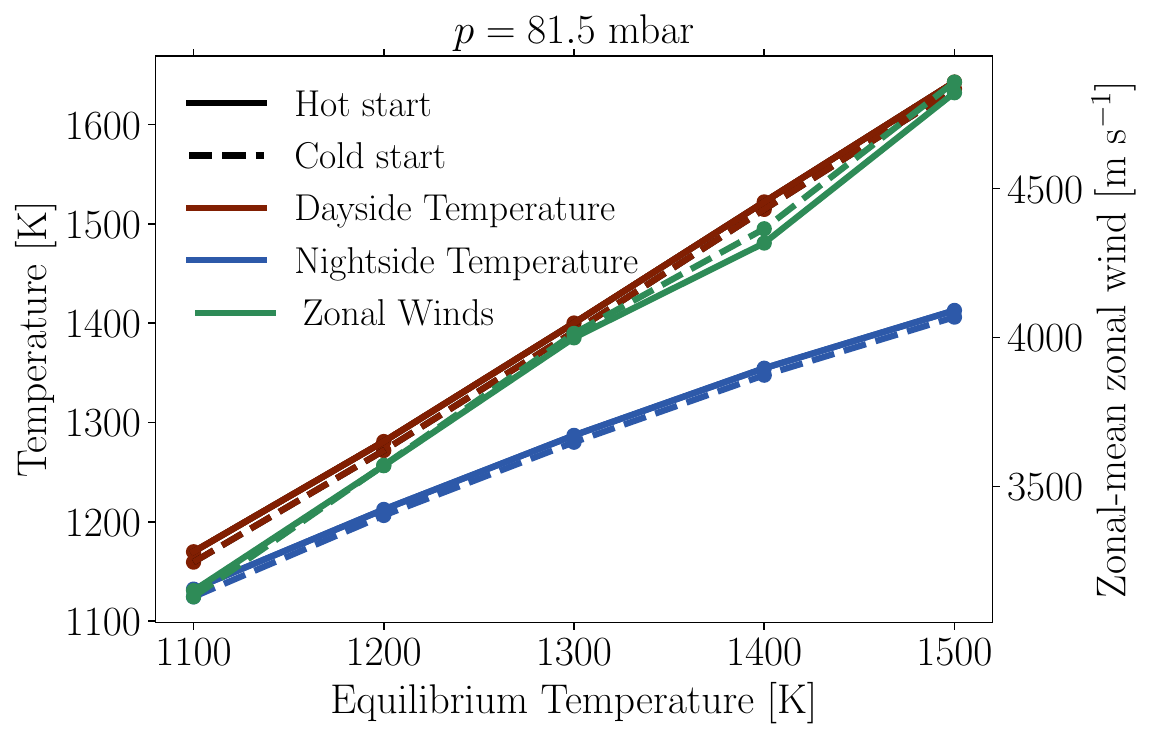}
    \caption{Dayside and nightside temperature as well as zonal-mean zonal wind speed at 81.5 mbar pressure as a function of equilibrium temperature from hot start cases (solid lines) and cold start cases (dashed lines) from the subset of cases in Suite 2(a). The dayside and nightside temperatures are averaged over the entire hemisphere, while the zonal-mean zonal wind is averaged within $10^\circ$ of the equator. While the temperatures in the hot start models are slightly higher than those in the cold start models due to the long radiative adjustment timescale of hot Jupiter atmospheres, I find no evidence for hysteresis in the averaged temperature or equatorial winds.}
    \label{fig:loop}
\end{figure}

In order to more quantitatively demonstrate this finding that the atmospheric dynamics of hot Jupiters exhibits limited dependence on initial thermal conditions, I show the dayside and nightside temperature and zonal-mean zonal winds from the cases with a hot initial thermal profile (``hot start,'' solid lines) and cold initial thermal profile (``cold start,'' dashed lines) for the subset of cases from in Suite 2(a) in \Fig{fig:loop}. 
If the simulations displayed a strong dependence on initial conditions, \Fig{fig:loop} would show %a hysteresis loop with distinct branches 
 significant differences between the cold start and hot start cases. Instead, I find that the near-photospheric temperature and zonal-mean zonal wind have a nearly equivalent dependence on equilibrium temperature regardless of the initial temperature profile. The small remaining differences between the cases are due to ongoing radiative adjustment towards equilibrium, causing the cold start cases to have slightly smaller day-night temperature differences and wind speeds than the hot start cases even after $5,000~\mathrm{days}$ of model time. However, there are no distinct branches for the cold and hot start cases in either temperature or winds. Additionally, as shown in Figures \ref{fig:tempwind} and \ref{fig:zonalwind}, there is no change in the temperature pattern or wind structure between the cold and hot start cases in Suite 2(a). These numerical results imply that the Rhines scale does not control the dynamical regime of hot Jupiters. This can be interpreted as a demonstration that the circulation of hot Jupiters is controlled by the large-scale day-to-night forcing, rather than an inverse turbulent energy cascade (N.T. Lewis, personal communication). 

\section{Analytic theory relating forcing to characteristic wind speeds}
\label{sec:theory}
%the potential forcing regimes in which an abrupt transition between equatorial sub-rotation and super-rotation is expected to occur in order to interpret my numerical results. 
Here I use an analytic approach broadly inspired by \cite{Shell:2004} to investigate the steady-state relationship between momentum forcing and the horizontal winds on hot Jupiters. To do so, I use a two-dimensional one and a half layer shallow-water model with day-to-night forcing prescribed as a Newtonian heating/cooling term from \cite{Perez-Becker:2013fv}. Here, the applied forcing (torque) is balanced by frictional drag, here assumed to be Rayleigh drag as in the numerical simulations in Suite 1 presented above, along with momentum exchange with the lower layer. Thus, the momentum forcing $F$ is equal to
\begin{equation}
    \label{eq:forcing}
    F = \frac{(h_\mathrm{eq} - h)}{\tau_\mathrm{rad}} \frac{u}{h} + \frac{u}{\tau_\mathrm{drag}}~\mathrm{,}
\end{equation}
where $h$ is the layer thickness, $h_\mathrm{eq}$ is the radiative equilibrium layer thickness, $\tau_\mathrm{rad}$ is the radiative timescale, $u$ is the zonal wind, and $\tau_\mathrm{drag} = k_F^{-1}$ is the drag timescale. Note that the day-to-night forcing term (first term on the right-hand side of Equation (10) and the frictional drag term (second term on the right-hand side of Equation (10)) must balance in a global average sense for the circulation to be steady state with $F = 0$. This is analogous to considering the global circulation as a heat engine \citep{Schubert:2013,Koll:2017}. As a result, one can consider $F$ as the ``residual'' forcing that can either increase or decrease wind speeds towards an equilibrium value.

I now use an approach similar to \cite{Perez-Becker:2013fv} to scale this expression to relate $F$ to the characteristic horizontal wind speed $u \sim U$ and the characteristic layer thickness $h \sim H$
\begin{equation}
    \label{eq:forcing}
    F \sim \frac{(\Delta h_\mathrm{eq} - \Delta h)}{\tau_\mathrm{rad}} \frac{U}{H} + \frac{U}{\tau_\mathrm{drag}} \approx \frac{(h_\mathrm{eq} - \Delta h)}{\tau_\mathrm{rad}} \frac{U}{H} + \frac{U}{\tau_\mathrm{drag}}~\mathrm{,}
\end{equation}
where $\Delta h$ is the characteristic day-to-night layer height contrast, $\Delta h_\mathrm{eq}$ is the characteristic day-to-night height contrast in radiative equilibrium, and I have assumed a large day-to-night radiative equilibrium height contrast $\Delta h_\mathrm{eq} \sim h_\mathrm{eq}$. %\bar{h}_\mathrm{eq}$.

Next, I use the scaled steady-state momentum balance from \cite{Perez-Becker:2013fv} to relate the day-night height contrast to planetary and model parameters:
\begin{equation}
    \label{eq:momentum}
    g \frac{\Delta h}{a} \sim \frac{U}{\tau_\mathrm{drag}} + fU~\mathrm{.}
\end{equation}
The left-hand side of \Eq{eq:momentum} is the pressure gradient driven by the day-to-night height contrast (where $a$ is the planetary radius), the first term on the right hand side is frictional drag, and the second term on the right hand side is the Coriolis force (where $f = 2 \Omega~\mathrm{sin}\phi$ is the Coriolis parameter). Solving for $\Delta h$, substituting it into \Eq{eq:forcing}, and collecting terms, I find a relationship between the momentum forcing $F$ and the characteristic horizontal winds $U$
\begin{equation}
    \label{equation:fu}
    \begin{split}
    F & \sim \frac{U}{H \tau_\mathrm{rad}} \left(h_\mathrm{eq} - \frac{a f U}{g} \right) + \frac{U}{\tau_\mathrm{drag}} \left(1 - \frac{aU}{g H \tau_\mathrm{rad}} \right) \\ 
    & \sim \frac{U}{\tau_\mathrm{rad}} \left(\frac{h_\mathrm{eq}}{H} - \frac{f U \tau_\mathrm{wave}^2}{a} \right) + \frac{U}{\tau_\mathrm{drag}}\left(1 - \frac{U \tau_\mathrm{wave}^2}{a \tau_\mathrm{rad}} \right) ~\mathrm{,}
    \end{split}
\end{equation}
where in the latter expression I have substituted $\tau_\mathrm{wave} \sim a/\sqrt{gH}$ \citep{Perez-Becker:2013fv}. Note that in the limit where $\tau_\mathrm{rad} \rightarrow \infty$, \Eq{equation:fu} reduces to $F \sim U/\tau_\mathrm{drag}$, the Rayleigh drag term. Similarly, in the limit where $\tau_\mathrm{drag} \rightarrow \infty$, \Eq{equation:fu} reduces to $F \sim \frac{h_\mathrm{eq}U}{H\tau_\mathrm{rad}} - \frac{f U^2 \tau_\mathrm{wave}^2}{a \tau_\mathrm{rad}}$ (i.e., only the first term on the right hand side is retained).
%to do, reduce using tau_wave?done 

\begin{figure*}
\centering
 \includegraphics[width=0.48\textwidth]{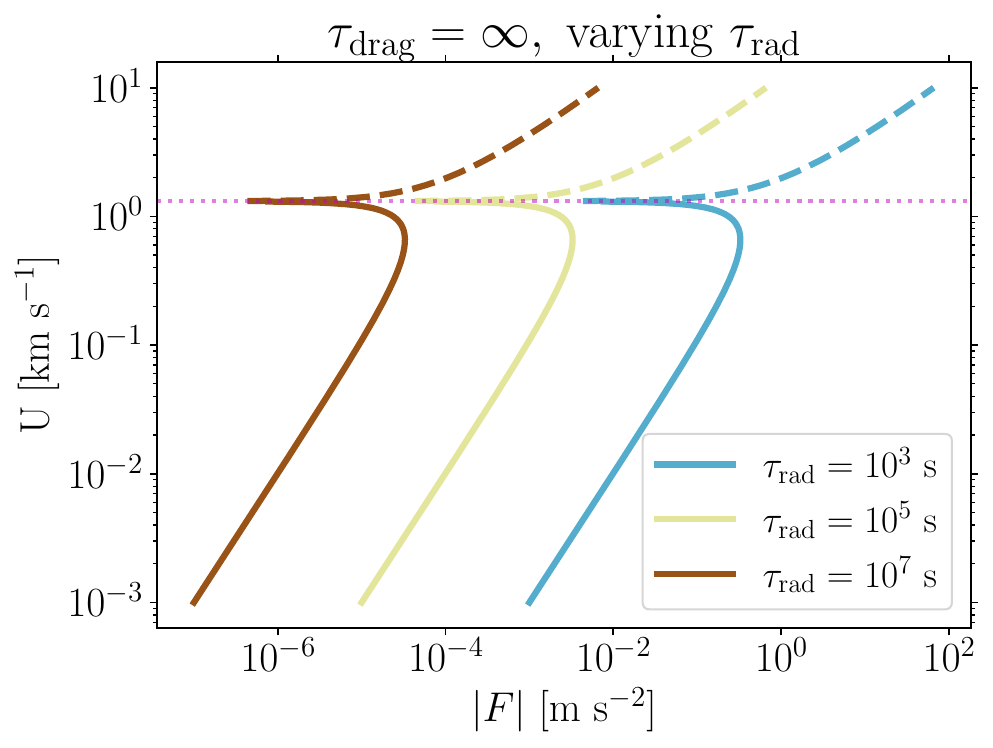}
  \includegraphics[width=0.48\textwidth]{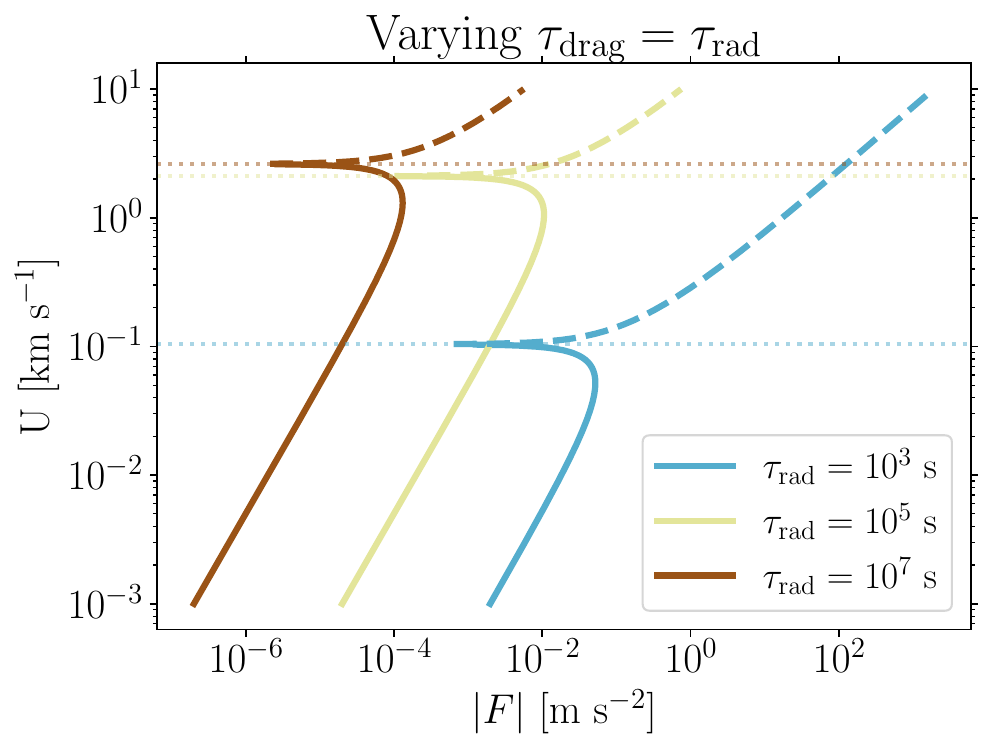}
    \caption{Relationship between the applied forcing $F$ and characteristic horizontal wind speed $U$ from Equation (\ref{equation:fu}) for cases with varying radiative timescale and no frictional drag ($\tau_\mathrm{drag} = \infty$, left hand side) and applied frictional drag with $\tau_\mathrm{drag} = \tau_\mathrm{rad}$ (right hand side). Planetary parameters correspond to the equator of HD 209458b. Colors correspond to varying radiative timescales. Solid lines correspond to positive values of $F$ and dashed lines correspond to negative values of $F$. Dotted horizontal lines display the equilibrium value of the characteristic wind speed (in the left hand plot, this value is the same for all $\tau_\mathrm{rad}$). I find that the forcing is positive at wind speeds below where $F$ has a minimum and negative at wind speeds above the minimum in $F$, implying that the horizontal winds only have one characteristic value at equilibrium for a given set of planetary parameters and combination of $\tau_\mathrm{drag}$ and $\tau_\mathrm{rad}$.}
    \label{fig:theoryplots}
\end{figure*}
\Fig{fig:theoryplots} shows the relationship between momentum forcing and characteristic horizontal wind speeds from \Eq{equation:fu} for varying $\tau_\mathrm{rad}$ and $\tau_\mathrm{drag}$ using planetary parameters ($a, f, g, T_\mathrm{eq}$) relevant for the equator of HD 209458b and assuming that $h_\mathrm{eq} \sim H \sim RT_\mathrm{eq}/g$. The local minimum in forcing corresponds to an equilibrium characteristic horizontal wind speed
\begin{equation}
\label{eq:ue}
    U_\mathrm{e} \sim \frac{a}{\tau_\mathrm{wave}^2} \left(\frac{\frac{h_\mathrm{eq}}{H} + \frac{\tau_\mathrm{rad}}{\tau_\mathrm{drag}}}{f + \frac{1}{\tau_\mathrm{drag}} }\right)~\mathrm{,}
\end{equation}
which is shown by the dotted lines in \Fig{fig:theoryplots}. Positive forcing (increasing the angular momentum) corresponds to characteristic horizontal wind speeds that are slower than the equilibrium value. Conversely, negative forcing corresponds to characteristic horizontal wind speeds that are faster than the equilibrium value in \Eq{eq:ue}.
%has a minimum at wind speeds of $\sim ~\mathrm{km}~\mathrm{s}^{-1}$ with weak drag,
%at a wind speed that depends on $\tau_\mathrm{drag}$ 
%and then becomes negative at faster wind speeds. 
Note that $U_e$ is independent of the radiative timescale $\tau_\mathrm{rad}$ in the case with $\tau_\mathrm{drag} = \infty$. This is because in the limit of weak drag ($\tau_\mathrm{drag} \rightarrow \infty$), \Eq{eq:ue} implies that 
%(in the strong forcing limit $h_\mathrm{eq} \sim H$) 
$U_\mathrm{e} \rightarrow \frac{h_\mathrm{eq} a}{H f \tau_\mathrm{wave}^2}$ %\sim \frac{RT}{af}$, 
independent of $\tau_\mathrm{rad}$. %Additionally, note that in the limit of strong drag ($\tau_\mathrm{drag} \rightarrow 0$), $U_\mathrm{e} \rightarrow 0$. 
%This implies that 
In general, an equilibrium wind speed can only be reached at the minimum in forcing, as at more positive values of $F$ (i.e., $U < U_e$) the forcing would drive wind speeds faster, toward equilibrium. Conversely, at more negative values of $F$ (i.e., $U > U_e$) the forcing would slow wind speeds, again toward their equilibrium value.
%at $U$ below the minimum in $F$ (i.e., $U < U_e$) the forcing is positive, driving faster $U$, while at $U$ above the minimum in $F$ (i.e., $U > U_e$) the forcing is negative, slowing winds. 
%This equilibrium wind speed depends on $\tau_\mathrm{drag}$, with shorter drag timescales leading to weaker winds in momentum balance, while it does not depend on $\tau_\mathrm{rad}$ because the momentum balance is not directly related to the radiative heating/cooling. 

%Overall, I find that the dependence of characteristic zonal wind speeds on forcing exhibits a single node, rather than a bifurcation in the wind speed with varying $U$ as in the Earth-like case with equator-to-pole forcing of \citep{Shell:2004}. This implies that the \textbf{characteristic wind} speeds of hot Jupiters are solely determined by planetary and atmospheric parameters and should not be dependent on initial conditions. 

The theory presented here is a linear scaling theory and thus does not incorporate non-linear feedbacks (e.g., hydrogen dissociation and recombination or other processes acting like moisture) that could amplify dependencies on initial conditions. This simplified theory notably does not explain why sub-Neptune GCMs \citep{Wang:2020aa} and rocky planet GCMs \citep{sergeev:2022ab} show bistability, while the hot Jupiter GCMs presented in this work and others (e.g., \citealp{Showman:2020rev}) do not. Further work is needed to incorporate the impacts of latent heat and latitude-dependent forcing to determine whether simple analytic theory along the lines of \cite{Shell:2004} can explain the transition between single jet and double jet regimes on tidally locked terrestrial planets. In addition, further work is needed to include the effects of hydrogen dissociation and recombination \citep{Komacek:2018aa,Roth:2021un} to explore whether the circulation of ultra-hot Jupiters (as opposed to cooler hot Jupiters) could exhibit hysteresis. 

%Implications for why sub-Neptune models \citep{Wang:2020aa} and rocky planet models \citep{sergeev:2022ab} show  bistability, while hot Jupiter GCMs in this work and others \citep{Showman:2020rev} do not. 
\section{Discussion}
\label{sec:disc}
\subsection{Limitations and future work}
\label{sec:limitations}
%Simplified RT, idealized models. 
This work directly builds upon the previous detailed study by \cite{Liu:2013} of the impact of initial conditions on the atmospheric dynamics of hot Jupiters though the following: first, incorporating a broader range of weak basal drag strengths in a suite of Newtonian cooling models; second, conducting a suite of simulations coupled to more realistic double-gray radiative transfer, and third, directly searching for a dependence of initial conditions on the atmospheric circulation at potential dynamical regime transitions. However, this work is limited in both model complexity and the range of initial conditions considered. Here I chose to use either Newtonian heating/cooling or double-gray radiative transfer in order to conduct a broad range of models, but these simulations are not as complex as those with more realistic non-gray radiative transfer. \cite{Lee:2021vo} demonstrated that non-gray radiative transfer (either using the picket fence or correlated-k approximations) provides a significant increase in model realism over double-gray approaches. Future work studying the potential for hysteresis in atmospheric circulation models should couple to non-gray radiative transfer, as is common in current state-of-the-art hot Jupiter GCMs \citep{Dobbs-Dixon:2013,Kataria:2014,Drummond:2018ab,Christie:2021tu,Parmentier:2021tt,Schneider:2022uu,Malsky:2023aa,Tan:2024aa}. In addition, the varied initial conditions in this work only covered the initial equatorial jet speed and temperature profile, and I did not consider different shear profiles for the initial velocity pattern. Furthermore, it would be more appropriate to search for hysteresis using simulations with an initial hot and cold condition that are then progressively adjusted to lower and higher values of instellation, respectively   (e.g., \citealp{2019ApJ...884L..46C,Turbet:2021aa}), as the simulations in this work simply test the potential for bifurcations rather than diagnosing a possible hysteresis loop.

The horizontal and vertical resolution in the simulations presented here is limited, with only 40 vertical layers and cubed-sphere horizontal resolutions of C16 and C32 for the two model suites presented. In this work, the resolution was kept low in order for these simulations to be integrated over a long model time (25,000 and 5,000 days in Suites 1 and 2 respectively) and reach steady-state. However, future work at a higher horizontal spatial resolution is required to more directly compare with predictions of circulation from high-resolution hot Jupiter GCMs with Newtonian cooling \citep{Cho:2021wb,2022MNRAS.511.3584S}, which display a much less steady circulation than previous low-resolution work coupled to double-gray radiative transfer \citep{Komacek:2020aa}. Longitudinal dissipation via numerical filters and hyperviscosity at low resolution has also been found to affect the resulting dynamics (e.g., \citealp{Heng:2011,Koll:2017,Christie:2024aa}), and any resulting hysteresis may be affected by the combination of horizontal resolution and applied numerical dissipation. In addition, simulations with a higher vertical resolution are required in order to determine if hot Jupiters undergo quasi-quadrennial oscillation-like stratospheric oscillations, as have been found in simulations of brown dwarfs and Jupiter with high vertical resolution \citep{Showman:2019us,Lian:2023aa}. Notably, simulations of temperate tidally locked planets exhibit quasi-biennial oscillation-like longitudinally asymmetric stratospheric oscillations \citep{Cohen:2022aa}, which could have a potential analogue in tidally locked gas giant simulations.

The impact of interior cooling on the deep atmospheric circulation was neglected in this work -- notably, in all cases with double-gray radiative transfer, the intrinsic temperature $T_\mathrm{int}$ (corresponding to the net thermal flux) was fixed to $100~\mathrm{K}$. Previous work by \cite{Carone:2019aa} found that the deep atmosphere can couple to the shallow atmosphere and drive subrotation on WASP-43b. In addition, \cite{Wang:2020aa} studied the atmospheric circulations of sub-Neptunes with thick atmospheres and found that their deep circulation could exhibit regime transitions in the dynamics over timescales of tens of thousands of days. Importantly, the interior evolution of hot Jupiters is strongly time-dependent, with deep heating applied over evolutionary timescales required to explain both the present-day radii of hot Jupiters \citep{Thorngren:2018,Sarkis:2021aa} as well as their re-inflation during the main-sequence lifetimes of their host stars \citep{Thorngren:2021aa}. As a result, the varying internal heat flux of hot Jupiters over time likely impacts their atmospheric circulation via the impact of interior heating on the three-dimensional radiative-convective boundary location \citep{Thorngren:2019aa,Zhang:2023ac}. Future work is required to study whether regime transitions in the atmospheric dynamics of hot Jupiters can occur as a function of interior heat flux.  

A variety of other model effects could also potentially lead to hysteresis in either the atmospheric chemistry or the climate state, for instance including disequilibrium chemistry via vertical and horizontal quenching \citep{Cooper:2006,Drummond:2018aa,Steinrueck:2018aa,Drummond:2020aa,Zamyatina:2022aa}, aerosol (cloud and haze) radiative feedback \citep{Christie:2021tu,Parmentier:2021tt,Roman:2021wl,Steinrueck:2021aa}, and molecular hydrogen dissociation and recombination \citep{Bell:2018aa,Tan:2019aa,Roth:2021un}. All of these relevant processes are not included here for simplicity, as instead in this work I have focused only on hysteresis driven by the dry dynamics. Importantly, the bistability found in the TRAPPIST-1e simulations of \cite{sergeev:2022ab} may be driven by moisture and the resulting impact of latent heating on the dynamics. Given that including hydrogen dissociation and recombination is analogous to including moisture (albeit in a non-dilute limit, \citealp{Bell:2018aa,Tan:2019aa}), it may be possible for ultra-hot Jupiters with high dayside atomic hydrogen mixing ratios to exhibit a dynamical bistability with regard to initial conditions. As a result, future work extending this study to the ultra-hot Jupiter regime in simulations that include the thermochemical effects of molecular hydrogen dissociation and recombination is required to determine the level to which molecular hydrogen dissociation and recombination can cause the atmospheric circulation of ultra-hot Jupiters to exhibit a dependence on initial conditions.

\subsection{Outlook}
The results presented here imply that the atmospheric circulation of hot Jupiters may exhibit a limited dependence on initial thermal forcing and wind profiles, in 
broad agreement with the broad similarity of results from hot Jupiter GCMs in the literature \citep{Heng:2014b,Showman:2020rev}. This implies that a wide range of GCM models will likely enable similar interpretation of observed phase curves and emission spectra.
%add wasp-43b paper citation and model ensemble connection here if paper published
These results further imply that population-level studies of hot Jupiters should be robust to large differences in thermal structure for a given set of planetary properties (e.g., equilibrium temperature, metallicity, gravity). The findings here also agree with the prediction of limited (but still potentially detectable) time-variability due to the lack of time-dependent transitions between model states \citep{Rauscher07,Dobbs-Dixon:2010aa,Komacek:2020aa}, which to date has been borne out in repeated Spitzer observations of hot Jupiter secondary eclipses and phase curves \citep{Agol:2010,Kilpatrick:2019aa,Murphy:2023aa}. 
%Note that this is distinct from the regime of more weakly-forced sub-Neptunes with deep atmospheres \citep{Wang:2020aa}, which may undergo .

Though there is qualitative agreement between many hot Jupiter modeling frameworks, there may still be important quantitative differences between model predictions. In this work, I only presented results from one model framework, using the MITgcm. There is a clear need to conduct model inter-comparisons for hot Jupiters, as has been done previously for temperate terrestrial planets \citep{Yang:2016aa,Yang:2019aa,2022PSJ.....3..213F,2022PSJ.....3..212S,2022PSJ.....3..211T}. Such an inter-comparison is currently being conducted through the MOCHA (MOdeling the Circulation of Hot exoplanet Atmospheres, led by N. Iro) exercise in the CUISINES framework \citep{2020GMD....13..707F}. Inter-comparisons like MOCHA are required in order to determine the extent to which conclusions presented in this work are valid in general, and to more broadly assess model agreement in the current era of JWST and improved high spectral resolution ground-based observations. 

\section{Conclusions}
\label{sec:conc}
In this work, I study the potential for a dependence of the atmospheric circulation of hot Jupiters on initial conditions, especially at dynamical regime boundaries where such hysteresis has been found in previous exoplanet atmospheric circulation models. To do so, I conduct two separate suites of GCM simulations, one varying the initial wind structure and one varying the initial temperature profile. Both of these suites of simulations show no significant dependence on initial conditions, implying that the atmospheric circulation of typical hot Jupiters may exhibit limited hysteresis. I summarize my findings as follows: 
\begin{enumerate}
    \item The suite of three-dimensional GCMs of hot Jupiter atmospheres with varying initial wind profile and basal drag strength show no strong dependence on their initial wind profile for both basal drag strengths considered. This result confirms the findings from the shallow water models and 3D GCMs of \cite{Liu:2013}, and extends them to cases with weaker basal drag strength and long model runtimes of $\sim 25,000$ Earth days.
    \item Hot Jupiters are expected from scaling arguments to potentially undergo a transition from intermediate rotators, where the Rossby deformation radius is smaller than the planetary radius but the Rhines scale is larger than the planetary radius, to rapid rotators, where both length scales are smaller than the planetary radius, with increasing instellation for a given host star type. Simulations of temperate rocky planets at another dynamical regime transition from intermediate to fast rotation have found a dependence of the resulting jet structure on the initial conditions \citep{sergeev:2022ab}. However, the hot Jupiter GCMs presented here display no transition in the circulation regime or clear hysteresis at this theoretically predicted regime boundary with varying initial temperature profile, unlike in the case of temperate rocky planets.
    \item These numerical results are interpreted in the context of a simple scaling theory leveraging the work of \cite{Perez-Becker:2013fv}. This theory demonstrates that that the characteristic horizontal day-to-night wind speed has a single equilibrium value as a function of the planetary-scale forcing of hot Jupiters, implying that the strong day-to-night forcing of hot Jupiters drives the insensitivity to initial conditions in numerical simulations.  
    %strong forcing of hot Jupiters drives the prevalence of superrotation in simulations alongside their insensitivity to initial conditions. 
    Though the atmospheric circulation of hot Jupiters displays at most limited hysteresis, hysteresis in the dynamics of temperate rocky planets and sub-Neptunes may be more likely due to the weaker large-scale forcing and the combination of day-to-night and equator-to-pole forcing contrasts that drives their circulation.
\end{enumerate}

\acknowledgments
 T.D.K. thanks the referee Neil Lewis for providing a deeply insightful review that significantly improved this manuscript. T.D.K. thanks Denis Sergeev, Jacob Haqq-Misra, and Arjun Savel for helpful comments on an early draft of this manuscript. T.D.K. also thanks Neil Lewis for insightful discussions on the utility of the Rhines Scale for tidally locked planets at Exoclimes VI. T.D.K. thanks Tiffany Kataria for guidance on setting up the MITgcm with varying initial wind profiles. T.D.K. acknowledges the University of Maryland supercomputing resources (\url{http://hpcc.umd.edu}) made available for conducting the research reported in this paper. The simulations with varying initial wind profiles were conducted on institutional computing resources at the University of Arizona Lunar and Planetary Laboratory, in collaboration with Adam Showman. T.D.K. further thanks Adam Showman for helpful discussions on the emergence of superrotation.

\bibliography{references,References_all,references_paste,References_terrestrial}

\begin{thebibliography}{}
\expandafter\ifx\csname natexlab\endcsname\relax\def\natexlab#1{#1}\fi
\providecommand{\url}[1]{\href{#1}{#1}}
\providecommand{\dodoi}[1]{doi:~\href{http://doi.org/#1}{\nolinkurl{#1}}}
\providecommand{\doeprint}[1]{\href{http://ascl.net/#1}{\nolinkurl{http://ascl.net/#1}}}
\providecommand{\doarXiv}[1]{\href{https://arxiv.org/abs/#1}{\nolinkurl{https://arxiv.org/abs/#1}}}

\bibitem[{Adcroft {et~al.}(2004)Adcroft, Hill, Campin, Marshall, \&
  Heimbach}]{Adcroft:2004}
Adcroft, A., Hill, C., Campin, J., Marshall, J., \& Heimbach, P. 2004, Monthly
  Weather Review, 132, 2845, \dodoi{10.1175/MWR2823.1}

\bibitem[{Agol {et~al.}(2010)Agol, Cowan, Knutson, Deming, Steffen, Henry, \&
  Charbonneau}]{Agol:2010}
Agol, E., Cowan, N., Knutson, H., {et~al.} 2010, The Astrophysical Journal,
  721, 1861, \dodoi{10.1088/0004-637X/721/2/1861}

\bibitem[{Batygin {et~al.}(2013)Batygin, Stanley, \& Stevenson}]{batygin_2013}
Batygin, K., Stanley, S., \& Stevenson, D. 2013, The Astrophysical Journal,
  776, 53, \dodoi{10.1088/0004-637X/776/1/53}

\bibitem[{Bell \& Cowan(2018)}]{Bell:2018aa}
Bell, T., \& Cowan, N. 2018, The Astrophysical Journal Letters, 857, L20,
  \dodoi{10.3847/2041-8213/aabcc8}

\bibitem[{{Bell} {et~al.}(2021){Bell}, {Dang}, {Cowan}, {Bean}, {D{\'e}sert},
  {Fortney}, {Keating}, {Kempton}, {Kreidberg}, {Line}, {Mansfield},
  {Parmentier}, {Stevenson}, {Swain}, \& {Zellem}}]{Bell:2021aa}
{Bell}, T.~J., {Dang}, L., {Cowan}, N.~B., {et~al.} 2021, \mnras, 504, 3316,
  \dodoi{10.1093/mnras/stab1027}

\bibitem[{{Bell} {et~al.}(2024){Bell}, {Crouzet}, {Cubillos}, {Kreidberg},
  {Piette}, {Roman}, {Barstow}, {Blecic}, {Carone}, {Coulombe}, {Ducrot},
  {Hammond}, {Mendon{\c{c}}a}, {Moses}, {Parmentier}, {Stevenson},
  {Teinturier}, {Zhang}, {Batalha}, {Bean}, {Benneke}, {Charnay}, {Chubb},
  {Demory}, {Gao}, {Lee}, {L{\'o}pez-Morales}, {Morello}, {Rauscher}, {Sing},
  {Tan}, {Venot}, {Wakeford}, {Aggarwal}, {Ahrer}, {Alam}, {Baeyens},
  {Barrado}, {Caceres}, {Carter}, {Casewell}, {Challener}, {Crossfield},
  {Decin}, {D{\'e}sert}, {Dobbs-Dixon}, {Dyrek}, {Espinoza}, {Feinstein},
  {Gibson}, {Harrington}, {Helling}, {Hu}, {Iro}, {Kempton}, {Kendrew},
  {Komacek}, {Krick}, {Lagage}, {Leconte}, {Lendl}, {Lewis}, {Lothringer},
  {Malsky}, {Mancini}, {Mansfield}, {Mayne}, {Mikal-Evans}, {Molaverdikhani},
  {Nikolov}, {Nixon}, {Palle}, {Petit dit de la Roche}, {Piaulet}, {Powell},
  {Rackham}, {Schneider}, {Steinrueck}, {Taylor}, {Welbanks}, {Yurchenko},
  {Zhang}, \& {Zieba}}]{Bell:2024aa}
{Bell}, T.~J., {Crouzet}, N., {Cubillos}, P.~E., {et~al.} 2024, Nature
  Astronomy, 8, 879, \dodoi{10.1038/s41550-024-02230-x}

\bibitem[{Beltz {et~al.}(2022)Beltz, Rauscher, Roman, \&
  Guilliat}]{Beltz:2022aa}
Beltz, H., Rauscher, E., Roman, M., \& Guilliat, A. 2022, The Astronomical
  Journal, 163, 35, \dodoi{10.3847/1538-3881/ac3746}

\bibitem[{Berardo {et~al.}(2017)Berardo, Cumming, \& Marleau}]{Berardo:2017aa}
Berardo, D., Cumming, A., \& Marleau, G. 2017, The Astrophysical Journal, 834,
  149, \dodoi{10.3847/1538-4357/834/2/149}

\bibitem[{{Carone} {et~al.}(2020){Carone}, {Baeyens}, {Molli{\`e}re}, {Barth},
  {Vazan}, {Decin}, {Sarkis}, {Venot}, \& {Henning}}]{Carone:2019aa}
{Carone}, L., {Baeyens}, R., {Molli{\`e}re}, P., {et~al.} 2020, Monthly Notices
  of the Royal Astronomical Society, 496, 3582, \dodoi{10.1093/mnras/staa1733}

\bibitem[{{Challener} \& {Rauscher}(2022)}]{Challener:2022}
{Challener}, R.~C., \& {Rauscher}, E. 2022, \aj, 163, 117,
  \dodoi{10.3847/1538-3881/ac4885}

\bibitem[{{Checlair} {et~al.}(2019){Checlair}, {Olson}, {Jansen}, \&
  {Abbot}}]{2019ApJ...884L..46C}
{Checlair}, J.~H., {Olson}, S.~L., {Jansen}, M.~F., \& {Abbot}, D.~S. 2019,
  \apjl, 884, L46, \dodoi{10.3847/2041-8213/ab487d}

\bibitem[{Cho {et~al.}(2015)Cho, Polichtchouk, \& Thrastarson}]{Cho:2015}
Cho, J., Polichtchouk, I., \& Thrastarson, H. 2015, Monthly Notices of the
  Royal Astronomical Society, 454, 3423, \dodoi{10.1093/mnras/stv1947}

\bibitem[{Cho {et~al.}(2021)Cho, Skinner, \& Thrastarson}]{Cho:2021wb}
Cho, J., Skinner, J., \& Thrastarson, H. 2021, The Astrophysical Journal
  Letters, 913, L32, \dodoi{10.3847/2041-8213/abfd37}

\bibitem[{Christie {et~al.}(2021)Christie, Mayne, Lines, \& {et
  al.}}]{Christie:2021tu}
Christie, D., Mayne, N., Lines, S., \& {et al.} 2021, Monthly Notices of the
  Royal Astronomical Society, 506, 4500, \dodoi{10.1093/mnras/stab2027}

\bibitem[{Christie {et~al.}(2024)Christie, Mayne, Zamyatina, Baskett,
  Mikal-Evans, Wood, \& Kohary}]{Christie:2024aa}
Christie, D., Mayne, N., Zamyatina, M., {et~al.} 2024, Monthly Notices of the
  Royal Astronomical Society, 532, 3001, \dodoi{10.1093/mnras/stae1408}

\bibitem[{{Cohen} {et~al.}(2022){Cohen}, {Bollasina}, {Palmer}, {Sergeev},
  {Boutle}, {Mayne}, \& {Manners}}]{Cohen:2022aa}
{Cohen}, M., {Bollasina}, M.~A., {Palmer}, P.~I., {et~al.} 2022, \apj, 930,
  152, \dodoi{10.3847/1538-4357/ac625d}

\bibitem[{Cooper \& Showman(2006)}]{Cooper:2006}
Cooper, C., \& Showman, A. 2006, The Astrophysical Journal, 649, 1048,
  \dodoi{10.1086/506312}

\bibitem[{{Coulombe} {et~al.}(2023){Coulombe}, {Benneke}, {Challener},
  {Piette}, {Wiser}, {Mansfield}, {MacDonald}, {Beltz}, {Feinstein}, {Radica},
  {Savel}, {Dos Santos}, {Bean}, {Parmentier}, {Wong}, {Rauscher}, {Komacek},
  {Kempton}, {Tan}, {Hammond}, {Lewis}, {Line}, {Lee}, {Shivkumar},
  {Crossfield}, {Nixon}, {Rackham}, {Wakeford}, {Welbanks}, {Zhang}, {Batalha},
  {Berta-Thompson}, {Changeat}, {D{\'e}sert}, {Espinoza}, {Goyal},
  {Harrington}, {Knutson}, {Kreidberg}, {L{\'o}pez-Morales}, {Shporer}, {Sing},
  {Stevenson}, {Aggarwal}, {Ahrer}, {Alam}, {Bell}, {Blecic}, {Caceres},
  {Carter}, {Casewell}, {Crouzet}, {Cubillos}, {Decin}, {Fortney}, {Gibson},
  {Heng}, {Henning}, {Iro}, {Kendrew}, {Lagage}, {Leconte}, {Lendl},
  {Lothringer}, {Mancini}, {Mikal-Evans}, {Molaverdikhani}, {Nikolov}, {Ohno},
  {Palle}, {Piaulet}, {Redfield}, {Roy}, {Tsai}, {Venot}, \&
  {Wheatley}}]{Coulombe:2023aa}
{Coulombe}, L.-P., {Benneke}, B., {Challener}, R., {et~al.} 2023, Nature, 620,
  292, \dodoi{10.1038/s41586-023-06230-1}

\bibitem[{Dang {et~al.}(2018)Dang, Cowan, Schwartz, \& {et al.}}]{Dang:2018aa}
Dang, L., Cowan, N., Schwartz, J., \& {et al.} 2018, Nature Astronomy, 2, 220,
  \dodoi{10.1038/s41550-017-0351-6}

\bibitem[{Dobbs-Dixon \& Agol(2013)}]{Dobbs-Dixon:2013}
Dobbs-Dixon, I., \& Agol, E. 2013, Monthly Notices of the Royal Astronomical
  Society, 435, 3159, \dodoi{10.1093/mnras/stt1509}

\bibitem[{Dobbs-Dixon {et~al.}(2010)Dobbs-Dixon, Cumming, \&
  Lin}]{Dobbs-Dixon:2010aa}
Dobbs-Dixon, I., Cumming, A., \& Lin, D. 2010, The Astrophysical Journal, 710,
  1395, \dodoi{10.1088/0004-637X/710/2/1395}

\bibitem[{Drummond {et~al.}(2018{\natexlab{a}})Drummond, Mayne, Manners, \& {et
  al.}}]{Drummond:2018ab}
Drummond, B., Mayne, N., Manners, J., \& {et al.} 2018{\natexlab{a}}, The
  Astrophysical Journal Letters, 855, L31, \dodoi{10.3847/2041-8213/aab209}

\bibitem[{Drummond {et~al.}(2018{\natexlab{b}})Drummond, Mayne, Manners, \& {et
  al.}}]{Drummond:2018aa}
---. 2018{\natexlab{b}}, The Astrophysical Journal, 869, 28,
  \dodoi{10.3847/1538-4357/aaeb28}

\bibitem[{Drummond {et~al.}(2020)Drummond, Hebrard, Mayne, Venot, r.J. Ridgway,
  Changeat, Tsai, Manners, Tremblin, Abraham, Sing, \&
  Kohary}]{Drummond:2020aa}
Drummond, B., Hebrard, E., Mayne, N., {et~al.} 2020, Astronomy {\&}
  Astrophysics, 636, A68, \dodoi{10.1051/0004-6361/201937153}

\bibitem[{{Fauchez} {et~al.}(2020){Fauchez}, {Turbet}, {Wolf}, {Boutle}, {Way},
  {Del Genio}, {Mayne}, {Tsigaridis}, {Kopparapu}, {Yang}, {Forget}, {Mandell},
  \& {Domagal Goldman}}]{2020GMD....13..707F}
{Fauchez}, T.~J., {Turbet}, M., {Wolf}, E.~T., {et~al.} 2020, Geoscientific
  Model Development, 13, 707, \dodoi{10.5194/gmd-13-707-2020}

\bibitem[{{Fauchez} {et~al.}(2022){Fauchez}, {Villanueva}, {Sergeev}, {Turbet},
  {Boutle}, {Tsigaridis}, {Way}, {Wolf}, {Domagal-Goldman}, {Forget},
  {Haqq-Misra}, {Kopparapu}, {Manners}, \& {Mayne}}]{2022PSJ.....3..213F}
{Fauchez}, T.~J., {Villanueva}, G.~L., {Sergeev}, D.~E., {et~al.} 2022, The
  Planetary Science Journal, 3, 213, \dodoi{10.3847/PSJ/ac6cf1}

\bibitem[{Guillot {et~al.}(1996)Guillot, Burrows, Hubbard, Lunine, \&
  Saumon}]{Guillot:1996}
Guillot, T., Burrows, A., Hubbard, W., Lunine, J., \& Saumon, D. 1996, The
  Astrophysical Journal Letters, 459, L35, \dodoi{10.1086/309935}

\bibitem[{Hammond \& Abbot(2022)}]{Hammond:2022aa}
Hammond, M., \& Abbot, D. 2022, Monthly Notices of the Royal Astronomical
  Society, 511, 2313, \dodoi{10.1093/mnras/stac228}

\bibitem[{Hammond \& Lewis(2021)}]{Hammond:2021aa}
Hammond, M., \& Lewis, N. 2021, Proceedings of the National Academy of Science,
  118, 2022705118, \dodoi{10.1073/pnas.2022705118}

\bibitem[{Hammond \& Pierrehumbert(2018)}]{Hammond:2018aa}
Hammond, M., \& Pierrehumbert, R. 2018, The Astrophysical Journal, 869, 65,
  \dodoi{10.3847/1538-4357/aaec03}

\bibitem[{{Hammond} {et~al.}(2020){Hammond}, {Tsai}, \&
  {Pierrehumbert}}]{Hammond:2020aa}
{Hammond}, M., {Tsai}, S.-M., \& {Pierrehumbert}, R.~T. 2020, \apj, 901, 78,
  \dodoi{10.3847/1538-4357/abb08b}

\bibitem[{{Hammond} {et~al.}(2024){Hammond}, {Bell}, {Challener}, {Lewis},
  {Weiner Mansfield}, {Malsky}, {Rauscher}, {Bean}, {Carone}, {Mendon{\c{c}}a},
  {Teinturier}, {Tan}, {Crouzet}, {Kreidberg}, {Morello}, {Parmentier},
  {Blecic}, {D{\'e}sert}, {Helling}, {Lagage}, {Molaverdikhani}, {Nixon},
  {Rackham}, \& {Yang}}]{Hammond:2024aa}
{Hammond}, M., {Bell}, T.~J., {Challener}, R.~C., {et~al.} 2024, The
  Astronomical Journal, 168, 4, \dodoi{10.3847/1538-3881/ad434d}

\bibitem[{Haqq-Misra {et~al.}(2018)Haqq-Misra, Wolf, Joshi, Zhang, \&
  Kopparapu}]{Haqq2018}
Haqq-Misra, J., Wolf, E., Joshi, M., Zhang, X., \& Kopparapu, R. 2018, The
  Astrophysical Journal, 852, 67, \dodoi{10.3847/1538-4357/aa9f1f}

\bibitem[{Heng {et~al.}(2011)Heng, Menou, \& Phillips}]{Heng:2011}
Heng, K., Menou, K., \& Phillips, P. 2011, Monthly Notices of the Royal
  Astronomical Society, 413, 2380, \dodoi{10.1111/j.1365-2966.2011.18315.x}

\bibitem[{Heng \& Showman(2015)}]{Heng:2014b}
Heng, K., \& Showman, A. 2015, Annual Reviews in Earth and Planetary Sciences,
  43, 509, \dodoi{10.1146/annurev-earth-060614-105146}

\bibitem[{Iro {et~al.}(2005)Iro, B\'{e}zard, \& Guillot}]{Iro:2005}
Iro, N., B\'{e}zard, B., \& Guillot, T. 2005, Astronomy and Astrophysics, 436,
  719, \dodoi{10.1051/0004-6361:20048344}

\bibitem[{Kataria {et~al.}(2015)Kataria, Showman, Fortney, Stevenson, Line,
  Kriedberg, Bean, \& Desert}]{Kataria:2014}
Kataria, T., Showman, A., Fortney, J., {et~al.} 2015, The Astrophysical
  Journal, 801, 86, \dodoi{10.1088/0004-637X/801/2/86}

\bibitem[{Kempton \& Rauscher(2012)}]{Kempton:2012vk}
Kempton, E., \& Rauscher, E. 2012, The Astrophysical Journal, 751, 117,
  \dodoi{10.1088/0004-637X/751/2/117}

\bibitem[{Kilpatrick {et~al.}(2020)Kilpatrick, Kataria, Lewis, Zellem, Henry,
  Cowan, de~Wit, Fortney, Knutson, Seager, Showman, \&
  Tucker}]{Kilpatrick:2019aa}
Kilpatrick, B., Kataria, T., Lewis, N., {et~al.} 2020, The Astronomical
  Journal, 159, 51, \dodoi{10.3847/1538-3881/ab6223}

\bibitem[{Koll \& Komacek(2018)}]{Koll:2017}
Koll, D., \& Komacek, T. 2018, The Astrophysical Journal, 853, 133,
  \dodoi{10.3847/1538-4357/aaa3de}

\bibitem[{Komacek \& Showman(2016)}]{Komacek:2015}
Komacek, T., \& Showman, A. 2016, The Astrophysical Journal, 821, 16,
  \dodoi{10.3847/0004-637X/821/1/16}

\bibitem[{Komacek \& Showman(2020)}]{Komacek:2020aa}
---. 2020, The Astrophysical Journal, 888, 2, \dodoi{10.3847/1538-4357/ab5b0b}

\bibitem[{Komacek {et~al.}(2017)Komacek, Showman, \& Tan}]{Komacek:2017}
Komacek, T., Showman, A., \& Tan, X. 2017, The Astrophysical Journal, 835, 198,
  \dodoi{10.3847/1538-4357/835/2/198}

\bibitem[{Komacek \& Tan(2018)}]{Komacek:2018aa}
Komacek, T., \& Tan, X. 2018, Research notes of the AAS, 2, 36,
  \dodoi{10.3847/2515-5172/aac5e7}

\bibitem[{Kylling \& Stamnes(1992)}]{Kylling:1992}
Kylling, A., \& Stamnes, K. 1992, Journal of Computational Physics, 102, 265,
  \dodoi{10.1016/0021-9991(92)90371-5}

\bibitem[{Laughlin {et~al.}(2011)Laughlin, Crismani, \& Adams}]{Laughlin_2011}
Laughlin, G., Crismani, M., \& Adams, F. 2011, The Astrophysical Journal
  Letters, 729, L7, \dodoi{10.1088/2041-8205/729/1/L7}

\bibitem[{Lee {et~al.}(2021)Lee, Parmentier, Hammond, Grimm, Kitzmann, Tan,
  Tsai, \& Pierrehumbert}]{Lee:2021vo}
Lee, E., Parmentier, V., Hammond, M., {et~al.} 2021, Monthly Notices of the
  Royal Astronomical Society, 506, 2695, \dodoi{10.1093/mnras/stab1851}

\bibitem[{Lewis \& Hammond(2022)}]{Lewis:2022aa}
Lewis, N., \& Hammond, M. 2022, The Astrophysical Journal, 941, 171,
  \dodoi{10.3847/1538-4357/ac8fed}

\bibitem[{{Lian} {et~al.}(2023){Lian}, {Tan}, \& {Hu}}]{Lian:2023aa}
{Lian}, Y., {Tan}, X., \& {Hu}, Y. 2023, \apj, 958, 50,
  \dodoi{10.3847/1538-4357/acfca6}

\bibitem[{Liu \& Showman(2013)}]{Liu:2013}
Liu, B., \& Showman, A. 2013, The Astrophysical Journal, 770, 42,
  \dodoi{10.1088/0004-637X/770/1/42}

\bibitem[{Lothringer {et~al.}(2018)Lothringer, Barman, \&
  Koskinen}]{Lothringer:2018aa}
Lothringer, J., Barman, T., \& Koskinen, T. 2018, The Astrophysical Journal,
  866, 27, \dodoi{10.3847/1538-4357/aadd9e}

\bibitem[{{Malsky} {et~al.}(2024){Malsky}, {Rauscher}, {Roman}, {Lee}, {Beltz},
  {Savel}, {Kempton}, \& {Cinque}}]{Malsky:2023aa}
{Malsky}, I., {Rauscher}, E., {Roman}, M.~T., {et~al.} 2024, The Astrophysical
  Journal, 961, 66, \dodoi{10.3847/1538-4357/ad0b70}

\bibitem[{Mansfield {et~al.}(2020)Mansfield, Schlawin, Lustig-Yaeger, Adams,
  Rauscher, Arcangeli, Feng, Gupta, Keating, Stevenson, \&
  Beatty}]{Mansfield:2020ab}
Mansfield, M., Schlawin, E., Lustig-Yaeger, J., {et~al.} 2020, Monthly Notices
  of the Royal Astronomical Society, 499, 5151, \dodoi{10.1093/mnras/staa3179}

\bibitem[{{May} {et~al.}(2022){May}, {Stevenson}, {Bean}, {Bell}, {Cowan},
  {Dang}, {Desert}, {Fortney}, {Keating}, {Kempton}, {Komacek}, {Lewis},
  {Mansfield}, {Morley}, {Parmentier}, {Rauscher}, {Swain}, {Zellem}, \&
  {Showman}}]{2022AJ....163..256M}
{May}, E.~M., {Stevenson}, K.~B., {Bean}, J.~L., {et~al.} 2022, \aj, 163, 256,
  \dodoi{10.3847/1538-3881/ac6261}

\bibitem[{Mikal-Evans {et~al.}(2023)Mikal-Evans, Sing, Dong, \& {et
  al.}}]{Mikal-Evans:2023aa}
Mikal-Evans, T., Sing, D., Dong, J., \& {et al.} 2023, The Astrophysical
  Journal Letters, 943, L17, \dodoi{10.3847/2041-8213/acb049}

\bibitem[{{Murphy} {et~al.}(2023){Murphy}, {Beatty}, {Roman}, {Malsky},
  {Wingate}, {Ochs}, {Cinque}, {Beltz}, {Rauscher}, {Kempton}, \&
  {Stevenson}}]{Murphy:2023aa}
{Murphy}, M.~M., {Beatty}, T.~G., {Roman}, M.~T., {et~al.} 2023, \aj, 165, 107,
  \dodoi{10.3847/1538-3881/acaec5}

\bibitem[{Noda {et~al.}(2017)Noda, Ishiwatari, Nakajima, Takahashi, Takehiro,
  Onishi, Hashimoto, Kuramoto, \& Hayashi}]{Noda:2017aa}
Noda, S., Ishiwatari, M., Nakajima, K., {et~al.} 2017, Icarus, 282, 1,
  \dodoi{10.1016/j.icarus.2016.09.004}

\bibitem[{Parmentier {et~al.}(2021)Parmentier, Showman, \&
  Fortney}]{Parmentier:2021tt}
Parmentier, V., Showman, A., \& Fortney, J. 2021, Monthly Notices of the Royal
  Astronomical Society, 501, 78, \dodoi{10.1093/mnras/staa3418}

\bibitem[{Perez-Becker \& Showman(2013)}]{Perez-Becker:2013fv}
Perez-Becker, D., \& Showman, A. 2013, The Astrophysical Journal, 776, 134,
  \dodoi{10.1088/0004-637X/776/2/134}

\bibitem[{Perna {et~al.}(2010)Perna, Menou, \& Rauscher}]{Perna_2010_1}
Perna, R., Menou, K., \& Rauscher, E. 2010, The Astrophysical Journal, 719,
  1421, \dodoi{10.1088/0004-637X/719/2/1421}

\bibitem[{Pierrehumbert \& Hammond(2019)}]{Pierrehumbert:2019vk}
Pierrehumbert, R., \& Hammond, M. 2019, Annual Review of Fluid Mechanics, 51,
  275, \dodoi{10.1146/annurev-fluid-010518-040516}

\bibitem[{{Rauscher} \& {Kempton}(2014)}]{Rauscher:2014aa}
{Rauscher}, E., \& {Kempton}, E. M.~R. 2014, \apj, 790, 79,
  \dodoi{10.1088/0004-637X/790/1/79}

\bibitem[{Rauscher \& Menou(2013)}]{Rauscher_2013}
Rauscher, E., \& Menou, K. 2013, The Astrophysical Journal, 764, 103,
  \dodoi{10.1088/0004-637X/764/1/103}

\bibitem[{{Rauscher} {et~al.}(2007){Rauscher}, {Menou}, {Cho}, {Seager}, \&
  {Hansen}}]{Rauscher07}
{Rauscher}, E., {Menou}, K., {Cho}, J.~Y.-K., {Seager}, S., \& {Hansen},
  B.~M.~S. 2007, The Astrophysical Journal Letters, 662, L115,
  \dodoi{10.1086/519374}

\bibitem[{Rhines(1975)}]{Rhines:1975aa}
Rhines, P. 1975, Journal of Fluid Mechanics, 69, 417,
  \dodoi{10.1017/S0022112075001504}

\bibitem[{Rogers(2017)}]{Rogers:2017}
Rogers, T. 2017, Nature Astronomy, 1, 131, \dodoi{10.1038/s41550-017-0131}

\bibitem[{Rogers \& Showman(2014)}]{Rogers:2020}
Rogers, T., \& Showman, A. 2014, The Astrophysical Journal Letters, 782, L4,
  \dodoi{10.1088/2041-8205/782/1/L4}

\bibitem[{Roman {et~al.}(2021)Roman, Kempton, Rauscher, Harada, Bean, \&
  Stevenson}]{Roman:2021wl}
Roman, M., Kempton, E., Rauscher, E., {et~al.} 2021, The Astrophysical Journal,
  908, 101, \dodoi{10.3847/1538-4357/abd549}

\bibitem[{Roth {et~al.}(2021)Roth, Drummond, Hebrard, Tremblin, Goyal, \&
  Mayne}]{Roth:2021un}
Roth, A., Drummond, B., Hebrard, E., {et~al.} 2021, Monthly Notices of the
  Royal Astronomical Society, 505, 4515, \dodoi{10.1093/mnras/stab1256}

\bibitem[{Roth {et~al.}(2024)Roth, Parmentier, \& Hammond}]{Roth:2024aa}
Roth, A., Parmentier, V., \& Hammond, M. 2024, Monthly Notices of the Royal
  Astronomical Society, 531, 1056, \dodoi{10.1093/mnras/stae984}

\bibitem[{Sarkis {et~al.}(2021)Sarkis, Mordasini, Henning, Marleau, \&
  Molli\'{e}re}]{Sarkis:2021aa}
Sarkis, P., Mordasini, C., Henning, T., Marleau, G., \& Molli\'{e}re, P. 2021,
  Astronomy {\&} Astrophysics, 645, A79, \dodoi{10.1051/0004-6361/202038361}

\bibitem[{Schneider {et~al.}(2022)Schneider, Carone, Decin, Jorgensen,
  Molli\'{e}re, Baeyens, Kiefer, \& Helling}]{Schneider:2022uu}
Schneider, A., Carone, L., Decin, L., {et~al.} 2022, Astronomy {\&}
  Astrophysics, 664, A56, \dodoi{10.1051/0004-6361/202142728}

\bibitem[{Schubert \& Mitchell(2013)}]{Schubert:2013}
Schubert, G., \& Mitchell, J. 2013, Comparative Climatology of Terrestrial
  Planets, ed. S.~Mackwell, A.~Simon-Miller, J.~Harder, \& M.~Bullock (Tucson,
  AZ: University of Arizona Press),
  \dodoi{10.2458/azu_uapress_9780816530595-ch008}

\bibitem[{Sergeev {et~al.}(2022)Sergeev, Lewis, Lambert, Mayne, Boutle,
  Manners, \& Kohary}]{sergeev:2022ab}
Sergeev, D., Lewis, N., Lambert, F., {et~al.} 2022, The Planetary Science
  Journal, 3, 214, \dodoi{10.3847/PSJ/ac83be}

\bibitem[{{Sergeev} {et~al.}(2022){Sergeev}, {Fauchez}, {Turbet}, {Boutle},
  {Tsigaridis}, {Way}, {Wolf}, {Domagal-Goldman}, {Forget}, {Haqq-Misra},
  {Kopparapu}, {Lambert}, {Manners}, \& {Mayne}}]{2022PSJ.....3..212S}
{Sergeev}, D.~E., {Fauchez}, T.~J., {Turbet}, M., {et~al.} 2022, The Planetary
  Science Journal, 3, 212, \dodoi{10.3847/PSJ/ac6cf2}

\bibitem[{Shell \& Held(2004)}]{Shell:2004}
Shell, K., \& Held, I. 2004, Journal of the Atmospheric Sciences, 61, 2928,
  \dodoi{10.1175/JAS-3312.1}

\bibitem[{Showman {et~al.}(2010)Showman, Cho, \& Menou}]{Showman_2009}
Showman, A., Cho, J., \& Menou, K. 2010, Exoplanets, ed. S.~Seager (Tucson, AZ:
  University of Arizona Press), \dodoi{10.48550/arXiv.0911.3170}

\bibitem[{Showman {et~al.}(2013)Showman, Fortney, Lewis, \&
  Shabram}]{showman_2013_doppler}
Showman, A., Fortney, J., Lewis, N., \& Shabram, M. 2013, The Astrophysical
  Journal, 762, 24, \dodoi{10.1088/0004-637X/762/1/24}

\bibitem[{Showman \& Guillot(2002)}]{showman_2002}
Showman, A., \& Guillot, T. 2002, Astronomy and Astrophysics, 385, 166,
  \dodoi{10.1051/0004-6361:20020101}

\bibitem[{Showman {et~al.}(2015)Showman, Lewis, \& Fortney}]{Showman:2014}
Showman, A., Lewis, N., \& Fortney, J. 2015, The Astrophysical Journal, 801,
  95, \dodoi{10.1088/0004-637X/801/2/95}

\bibitem[{Showman \& Polvani(2010)}]{Showman:2010}
Showman, A., \& Polvani, L. 2010, Geophysical Research Letters, 37, L18811,
  \dodoi{10.1029/2010GL044343}

\bibitem[{Showman \& Polvani(2011)}]{Showman_Polvani_2011}
---. 2011, The Astrophysical Journal, 738, 71,
  \dodoi{10.1088/0004-637X/738/1/71}

\bibitem[{Showman {et~al.}(2020)Showman, Tan, \& Parmentier}]{Showman:2020rev}
Showman, A., Tan, X., \& Parmentier, V. 2020, Space Science Reviews, 216, 139,
  \dodoi{10.1007/s11214-020-00758-8}

\bibitem[{Showman {et~al.}(2019)Showman, Tan, \& Zhang}]{Showman:2019us}
Showman, A., Tan, X., \& Zhang, X. 2019, The Astrophysical Journal, 883, 4,
  \dodoi{10.3847/1538-4357/ab384a}

\bibitem[{{Skinner} \& {Cho}(2021)}]{2021MNRAS.504.5172S}
{Skinner}, J.~W., \& {Cho}, J.~Y.~K. 2021, \mnras, 504, 5172,
  \dodoi{10.1093/mnras/stab971}

\bibitem[{{Skinner} \& {Cho}(2022)}]{2022MNRAS.511.3584S}
---. 2022, \mnras, 511, 3584, \dodoi{10.1093/mnras/stab2809}

\bibitem[{Stamnes {et~al.}(1988)Stamnes, Tsay, Wiscombe, \&
  Jayaweera}]{Stamnes:2027}
Stamnes, K., Tsay, S., Wiscombe, W., \& Jayaweera, K. 1988, Applied Optics, 27,
  2502, \dodoi{10.1364/AO.27.002502}

\bibitem[{Steinrueck {et~al.}(2019)Steinrueck, Parmentier, Showman, Lothringer,
  \& Lupu}]{Steinrueck:2018aa}
Steinrueck, M., Parmentier, V., Showman, A., Lothringer, J., \& Lupu, R. 2019,
  The Astrophysical Journal, 880, 14, \dodoi{10.3847/1538-4357/ab2598}

\bibitem[{Steinrueck {et~al.}(2021)Steinrueck, Showman, Lavvas, Koskinen, Tan,
  \& Zhang}]{Steinrueck:2021aa}
Steinrueck, M., Showman, A., Lavvas, P., {et~al.} 2021, Monthly Notices of the
  Royal Astronomical Society, 504, 2783, \dodoi{10.1093/mnras/stab1053}

\bibitem[{Tan \& Komacek(2019)}]{Tan:2019aa}
Tan, X., \& Komacek, T. 2019, The Astrophysical Journal, 886, 26,
  \dodoi{10.3847/1538-4357/ab4a76}

\bibitem[{{Tan} {et~al.}(2024){Tan}, {Komacek}, {Batalha}, {Deming}, {Lupu},
  {Parmentier}, \& {Pierrehumbert}}]{Tan:2024aa}
{Tan}, X., {Komacek}, T.~D., {Batalha}, N.~E., {et~al.} 2024, \mnras,
  \dodoi{10.1093/mnras/stae050}

\bibitem[{Thorngren \& Fortney(2018)}]{Thorngren:2018}
Thorngren, D., \& Fortney, J. 2018, The Astronomical Journal, 155, 214,
  \dodoi{10.3847/1538-3881/aaba13}

\bibitem[{Thorngren {et~al.}(2021)Thorngren, Fortney, Lopez, Berger, \&
  Huber}]{Thorngren:2021aa}
Thorngren, D., Fortney, J., Lopez, E., Berger, T., \& Huber, D. 2021, The
  Astrophysical Journal Letters, 909, L16, \dodoi{10.3847/2041-8213/abe86d}

\bibitem[{Thorngren {et~al.}(2019)Thorngren, Gao, \&
  Fortney}]{Thorngren:2019aa}
Thorngren, D., Gao, P., \& Fortney, J. 2019, The Astrophysical Journal Letters,
  884, L6, \dodoi{10.3847/2041-8213/ab43d0}

\bibitem[{Tsai {et~al.}(2014)Tsai, Dobbs-Dixon, \& Gu}]{Tsai:2014}
Tsai, S., Dobbs-Dixon, I., \& Gu, P. 2014, The Astrophysical Journal, 793, 141,
  \dodoi{10.1088/0004-637X/793/2/141}

\bibitem[{{Turbet} {et~al.}(2021){Turbet}, {Bolmont}, {Chaverot}, {Ehrenreich},
  {Leconte}, \& {Marcq}}]{Turbet:2021aa}
{Turbet}, M., {Bolmont}, E., {Chaverot}, G., {et~al.} 2021, \nat, 598, 276,
  \dodoi{10.1038/s41586-021-03873-w}

\bibitem[{{Turbet} {et~al.}(2022){Turbet}, {Fauchez}, {Sergeev}, {Boutle},
  {Tsigaridis}, {Way}, {Wolf}, {Domagal-Goldman}, {Forget}, {Haqq-Misra},
  {Kopparapu}, {Lambert}, {Manners}, {Mayne}, \& {Sohl}}]{2022PSJ.....3..211T}
{Turbet}, M., {Fauchez}, T.~J., {Sergeev}, D.~E., {et~al.} 2022, The Planetary
  Science Journal, 3, 211, \dodoi{10.3847/PSJ/ac6cf0}

\bibitem[{Wang \& Wordsworth(2020)}]{Wang:2020aa}
Wang, H., \& Wordsworth, R. 2020, The Astrophysical Journal, 891, 7,
  \dodoi{10.3847/1538-4357/ab6dcc}

\bibitem[{Wang {et~al.}(2018)Wang, Read, Tabataba-Vakili, \& Young}]{Wang:2018}
Wang, Y., Read, P., Tabataba-Vakili, F., \& Young, R. 2018, Quarterly Journal
  of the Royal Meteorological Society, 144, 2537, \dodoi{10.1002/qj.3350}

\bibitem[{{Yang} {et~al.}(2019){Yang}, {Leconte}, {Wolf}, {Merlis}, {Koll},
  {Forget}, \& {Abbot}}]{Yang:2019aa}
{Yang}, J., {Leconte}, J., {Wolf}, E.~T., {et~al.} 2019, \apj, 875, 46,
  \dodoi{10.3847/1538-4357/ab09f1}

\bibitem[{{Yang} {et~al.}(2016){Yang}, {Leconte}, {Wolf}, {Goldblatt}, {Feldl},
  {Merlis}, {Wang}, {Koll}, {Ding}, {Forget}, \& {Abbot}}]{Yang:2016aa}
---. 2016, \apj, 826, 222, \dodoi{10.3847/0004-637X/826/2/222}

\bibitem[{Youdin \& Zhu(2025)}]{Youdin:2025aa}
Youdin, A., \& Zhu, Z. 2025, arXiv e-prints:2501.13214v1

\bibitem[{Zamyatina {et~al.}(2023)Zamyatina, Hebrard, Drummond, Mayne, Manners,
  Christie, Tremblin, Sing, \& Kohary}]{Zamyatina:2022aa}
Zamyatina, M., Hebrard, E., Drummond, B., {et~al.} 2023, MNRAS, 519, 3129,
  \dodoi{10.1093/mnras/stac3432}

\bibitem[{Zhang {et~al.}(2018)Zhang, Knutson, Kataria, Schwartz, Cowan,
  Showman, Burrows, Fortney, Todorov, Desert, Agol, \& Deming}]{Zhang:2017a}
Zhang, M., Knutson, H., Kataria, T., {et~al.} 2018, The Astronomical Journal,
  155, 83, \dodoi{10.3847/1538-3881/aaa458}

\bibitem[{Zhang(2020)}]{Zhang:2020rev}
Zhang, X. 2020, Research in Astronomy and Astrophysics, 20, 099,
  \dodoi{10.1088/1674-4527/20/7/99}

\bibitem[{{Zhang} {et~al.}(2023){Zhang}, {Li}, {Ge}, \& {Le}}]{Zhang:2023ac}
{Zhang}, X., {Li}, C., {Ge}, H., \& {Le}, T. 2023, The Astrophysical Journal,
  957, 22, \dodoi{10.3847/1538-4357/acee7d}

\end{thebibliography}

%\clearpage
%\appendix
%{\bf 
%\section{Varying initial thermal structure near the intermediate-rapid rotator transition}
%}
%\subsection{Comparing hot and cold interior: pressure dependent-flow}

%\subsection{Varying irradiation and internal temperature}

%\subsection{Enhanced surface gravity}

%\vspace{-2em}
%\subsection{Methods}
%Simulation Setup and Analysis}

% \section{Author publication charges} \label{sec:pubcharge}

% The current cost for the different quanta types is available at 
% \url{https://journals.aas.org/article-charges-and-copyright/#author_publication_charges}. 
% Authors may use the ApJL length calculator to get a {\tt rough} estimate of 
% the number of word and float quanta in their manuscript. The calculator 
% is located at \url{https://authortools.aas.org/ApJL/betacountwords.html}.


\end{document}